\documentclass[aps,prd,amsmath,amssymb,preprintnumbers,preprint,nofootinbib,a4paper,11pt]{revtex4-1}
\pdfoutput=1
\usepackage{graphicx}
\graphicspath{{./myfigures/}}
\usepackage{url}
\usepackage[bookmarks, pagebackref=false]{hyperref}
\usepackage{color}
	\definecolor{rossoCP3}{cmyk}{0,.88,.77,.40}
		\definecolor{graa}{rgb}{0.8,0.8,0.8}
		\definecolor{blaa}{rgb}{0.2,0.2,0.6}
		\hypersetup{
			colorlinks, 
			bookmarksopen, 
			bookmarksnumbered,
			citecolor=blaa, 		
			linkcolor=rossoCP3,	
			urlcolor=rossoCP3,			
			}
\usepackage{amsthm}
\usepackage{bm}
\usepackage{bbm}
\usepackage{pxfonts}
\usepackage{youngtab}

\usepackage{placeins}
\usepackage{xspace}
\usepackage{cancel} 

\usepackage{slashed}
\usepackage[caption=false, labelformat=simple, listofformat=subsimple, labelfont=bf, margin=5pt, justification=raggedright]{subfig}

	\makeatletter
		\renewcommand{\p@subfigure}{}
	\makeatother

\usepackage{natbib}
\usepackage{array,multirow}
\usepackage{float}

\newcommand{\beq}{\begin{eqnarray}}
\newcommand{\eeq}{\end{eqnarray}}

\newcommand{\bmp}{\noindent\begin{minipage}{16cm}}
\newcommand{\emp}{\end{minipage}\vskip 7mm} 

\newcommand{\atheorem}{\emph{\lowercase{a}}~theorem\xspace}

\newcommand{\betafunction}{$\beta$~function\xspace}
\newcommand{\betafunctions}{$\beta$~functions\xspace}

\newcommand{\afree}{a_\textrm{free}}
\newcommand{\cfree}{c_\textrm{free}}
\newcommand{\Tr}{\text{Tr}}
\newcommand{\vev}[1]{\langle #1 \rangle}

\def\lsim{\mathrel{\rlap{\lower4pt\hbox{\hskip1pt$\sim$}}
    \raise1pt\hbox{$<$}}}                
\def\gsim{\mathrel{\rlap{\lower4pt\hbox{\hskip1pt$\sim$}}
    \raise1pt\hbox{$>$}}}                

\newcommand{\drawsquare}[2]{\hbox{%
\rule{#2pt}{#1pt}\hskip-#2pt
\rule{#1pt}{#2pt}\hskip-#1pt
\rule[#1pt]{#1pt}{#2pt}}\rule[#1pt]{#2pt}{#2pt}\hskip-#2pt
\rule{#2pt}{#1pt}}

\newcommand{\Yfund}{\raisebox{-.5pt}{\drawsquare{6.5}{0.4}}}
\newcommand{\Ysymm}{\raisebox{-.5pt}{\drawsquare{6.5}{0.4}}\hskip-0.4pt%
        \raisebox{-.5pt}{\drawsquare{6.5}{0.4}}}
\newcommand{\Yasymm}{\raisebox{-3.5pt}{\drawsquare{6.5}{0.4}}\hskip-6.9pt%
        \raisebox{3pt}{\drawsquare{6.5}{0.4}}}

\usepackage{color}
\definecolor{gray}{rgb}{0.5,0.5,0.5}

\begin{document}
\title{\texorpdfstring{\Large\color{rossoCP3} Conformal Data  of Fundamental Gauge-Yukawa Theories}{ The \atheorem for Gauge--Yukawa theories beyond Banks--Zaks}}
\author{Nicola Andrea Dondi}
\email{dondi@cp3.sdu.dk} 
\author{Francesco Sannino}
\email{sannino@cp3.dias.sdu.dk}  
 \affiliation{
{ \color{rossoCP3}  \rm CP}$^{\color{rossoCP3} \bf 3}${\color{rossoCP3}\rm--Origins} \& the Danish Institute for Advanced Study {\color{rossoCP3} \rm DIAS},\\ 
University of Southern Denmark, Campusvej 55, DK--5230 Odense M, Denmark.
} 
\author{Vladimir Prochazka}
\email{vladimir.prochazka@physics.uu.se} 
\affiliation{Department of Physics and Astronomy, Uppsala University,\\
Box 516, SE-75120 Uppsala, Sweden }
\begin{abstract}
 
 We determine central charges, critical exponents and appropriate gradient flow relations for nonsupersymmetric vector-like and chiral Gauge-Yukawa theories that are fundamental according to Wilson and that feature calculable UV or IR interacting fixed points.  We further uncover relations and identities among the various local and global conformal data. This information is used to provide the first extensive characterisation of general classes of free and safe quantum field theories of either chiral or vector-like nature via their conformal data. Using large $N_f$ techniques we also provide examples in which the safe fixed point is nonperturbative but for which conformal perturbation theory can be used to determine the global variation of the $a$ central charge.  

\vspace{0.5cm}
\noindent
{ \footnotesize  \it Preprint: CP$^3$-Origins-2017-057 DNRF90}
\end{abstract}

\maketitle

{\hypersetup{linkcolor=black}
\tableofcontents
}

\newpage

\section{Introduction}
The Standard Model is embodied by a gauge-Yukawa theory and constitutes one of the most successful theories of nature. It is therefore essential to deepen our understanding of these theories.

An important class of gauge-Yukawa theories is the one that, according to Wilson~\cite{Wilson:1971bg,Wilson:1971dh}, can be defined fundamental. This means that the theories belonging to this class are valid at arbitrary short and long distances. In practice this is ensured by requiring that a conformal field theory controls the short distance behaviour.  
Asymptotically free theories are a time-honoured example    
  \cite{Gross:1973ju,Politzer:1973fx} in which the ultraviolet is controlled by a not interacting conformal field theory. Another possibility is that an interacting ultraviolet fixed point emerges, these theories are known as asymptotically safe.  The first  proof of existence of asymptotically safe gauge-Yukawa theories in four dimensions appeared in  \cite{Litim:2014uca}. The original model has since enjoyed various extensions by inclusion of semi-simple gauge groups \cite{Bond:2017lnq} and supersymmetry \cite{Bond:2017suy, Bajc:2017xwx}. These type of theories constitute now an important alternative to asymptotic freedom.  One can now imagine new extensions of the Standard Model  \cite{Abel:2017ujy,Abel:2017rwl,Pelaggi:2017wzr,Mann:2017wzh,Pelaggi:2017abg,Bond:2017wut} and novel ways to achieve radiative symmetry breaking \cite{Abel:2017ujy,Abel:2017rwl}. In fact even QCD and QCD-like theories at large number of flavours can be argued to become safe \cite{Antipin:2017ebo,Pica:2010xq} leading to a novel testable {\it safe} revolution of the original QCD conformal window \cite{Sannino:2004qp,Dietrich:2006cm}.

The purpose of this paper is, at first, to determine the conformal data for generic gauge-Yukawa theories within perturbation theory.  We shall use the acquired information to relate various interesting quantities characterising the given conformal field theory.  We will then specialise our findings to determine the conformal data of several fundamental gauge-Yukawa theories at IR and UV interacting fixed points. 

The work is organised as follows: In Section~\ref{GYanalysis} we setup the notation, introduce the most general gauge-Yukawa theory and briefly summarise the needed building blocks to then in Sections~\ref{Local}  and \ref{Global} determine the explicit expressions in perturbation theory respectively for the local and global conformal data. We then specialise the results to the case of a gauge theory with a single Yukawa coupling in Section~\ref{sec:1Yukawa} because several models are of this type and because it helps to elucidate some of the salient points of our results. The relevant templates of asymptotically free and safe field theories are investigated in Section~\ref{templates}.  We offer our conclusions in Section~\ref{conclusions}. In the appendix we provide further technical details  and setup notation for the perturbative version of the a-theorem and  conformal perturbation theory. 
\section{Gauge-Yukawa theories}
\label{GYanalysis}

The classes of theories we are interested in can be described by the following Lagrangian template:
\begin{equation}
	\mathcal{L} = -\frac{1}{4g^2_a} F_{\mu\nu,a} F_a^{\mu\nu}
		+ i \, \Psi_i^\dag \bar{\sigma}^\mu D_\mu \Psi_i
		+ \frac{1}{2} D_\mu \phi_A D^\mu \phi_A
		- \left( y^A_{ij} \, \Psi_i \Psi_j \phi_A + \textrm{h.c.} \right)
		- \frac{1}{4!} \lambda_{ABCD} \, \phi_A \phi_B \phi_C \phi_D \ ,
	\label{eq:Lagrangian}
\end{equation}
where we dropped the gauge indices for $F_{\mu\nu,a}$, the Weyl fermions $\Psi_i$ the real scalar $\phi_A$ and the Yukawa and scalar coupling matrices. Differently from the notation in \cite{Jack:2014pua} \cite{Antipin:2013pya}, we make explicit the "flavour" indices, as we have in mind application to models containing fields in different gauge group representation. 
The index $a$ runs over the distinct gauge interactions constituting the semi-simple gauge group $\mathcal{G} = \otimes_a \mathcal{G}_a$. The  fermions and the scalar transform according to given representations $\mathcal{R}_{\psi_i}^a$  and $\mathcal{R}_{\phi_A}^a$ of the underlying simple gauge groups. The Yukawa and scalar coupling structures are such to respect gauge invariance.\\
 In certain cases it is convenient to separate the gauge-flavour structure of the Yukawa coupling from the coupling itself:  
   \begin{equation}
   y_{ij}^A \equiv \sum_I y^I T_{ij}^{IA} \ ,
   \end{equation}
  where $T_{ij}^{IA}$ is a coupling-free matrix and the indices $I,j$ run on all remaining indices. If flavour symmetries are present, the $T$ matrix will be such to preserve the symmetries. In this notation the individual  beta functions to the two-loop order for the gauge coupling and one-loop for the Yukawas read 
\begin{eqnarray}
	\beta_g^a & = & - \frac{g^3_a}{(4\pi)^2} \left[b_0^a + \frac{(b_1)^{ab}}{(4\pi)^2} g^2_b + \frac{(b_{y})_{IJ}^a }{(4\pi)^2} y^I y^J  \right] \ ,
	\label{eq:betagaugeM} \\
	\beta_{y}^I & = &\frac{1}{(4\pi)^2} \left[ (c_1)_{JKL}^I y^J y^K y^L + (c_{2})_{J}^{b I} g^2_b y^J \right] \ ,
	\label{eq:betaYukawaM} 
\end{eqnarray}
where repeated indices (except $a$ in \eqref{eq:betagaugeM})  are summed over. The above beta functions are general  with the coefficients depending on the specific underlying gauge-Yukawa theory. Furthermore, $(c_1)_{JKL}^I$ is totally symmetric in the last three (lower) indices as well as $(b_y)^a_{IJ}$ in the Yukawa ones.  To this order the scalar couplings do not run yet \cite{Antipin:2013sga}. Therefore to the present order, that we will refer as the 2-1-0, the $d=4$  Zamolodchikov two-index symmetric metric $\chi$ (c.f.\eqref{eq:consistencycondition}) over the couplings is fully diagonal and reads
\begin{equation}
	\chi =
	\left(\begin{array}{ccc}
		\frac{\chi_{g_a g_a}}{g_a^2} (1+ \frac{A_a}{(4\pi)^2}g_a^2)& 0 \\
		0 &  \chi_{y_Iy_I}
	\end{array}\right)  \ ,
\label{eq:metricM}
\end{equation}
where $\chi_{gg}, \; \chi_{y_Iy_I}$ and the $A_a$ quantities are coupling-independent constants.  Following Ref~\cite{Osborn:1989td} in order to prove that  the $\chi_{y_I y_J}$ part of the metric is diagonal we consider the corresponding operators $O_I=\Psi_1 \Psi_2 \phi_A  + (\text{h.c}) \ ,  $ and $O_J= \Psi_3 \Psi_4 \phi_B + (\text{h.c})$ and build the lowest order two-point function \begin{equation}
\vev{O_I O_J}_{\text{LO}} \sim \delta^{AB}( \delta_{13}\delta_{24} + \delta_{23}\delta_{14}) \ ,
\end{equation} 
which vanishes unless $O_I=O_J$ (I.e. when all the indices coincide). Off-diagonal terms will appear at higher orders.  Technically the $\vev{O_I O_J}$ leading perturbative contribution is a two-loop diagram justifying in \eqref{eq:metricM}  the inclusion of the two-loop gauge $A$-term. 
Explicitly, in our notation
\begin{equation}
	\chi_{g_a g_a} =  \frac{1}{(4\pi)^2}\frac{d(G_a)}{2},
	\quad
	\chi_{y_I y_I} = \frac{1}{(4\pi)^4} \frac{1}{6} 
		\sum_{A,i,j} \text{Tr}_g \left[  (T^{IA})_{ij} (T^{IA*})_{ij} \right]  \ , \quad A_a = 17C(\mathcal{G}_a) -\frac{10}{3}\sum_i T_{R^a_{\psi_i}} -  \frac{7}{6}\sum_i T_{R^a_{\phi_i}} \ .
	\label{eq:chis2l}
\end{equation}
The factor of $\frac{1}{(4\pi)^4}$ in $\chi_{y_I y_I}$ agrees with its two-loop nature.  \\
The Weyl consistency conditions, shown to be relevant also for Standard Model computations in \cite{Antipin:2013pya} and further tested and discussed in \cite{Bond:2017tbw, Jack:2016utw, Bednyakov:2015ooa, Jack:2015tka, Jack:2014pua, Molgaard:2014hpa}, are briefly reviewed in Appendix~\ref{3loops} (c.f. \eqref{eq:integrabilitycondition} and \eqref{eq:integCondConst} in particular) for the present system and yield the following scheme-independent relations among the gauge and Yukawa coefficients in the beta functions:
\begin{equation}\begin{split}
	\frac{1}{(4 \pi)^2} \chi_{g_a g_a} (b_{y})^a_{ IJ}  &= - \chi_{y_Iy_I} (c_{2})^{a I}_{J} \\
	\chi_{y_I y_I}  (c_1)_{JLK}^I &= \chi_{y_J y_J}(c_1)_{IKL}^J \\
	(c_2)^{a I}_{J}\, \chi_{y_I y_I} &= (c_2)^{a J}_{I}\, \chi_{y_J y_J}\\
	(b_1)^{ab}\, \chi_{g_a g_a} &= (b_1)^{ba}\, \chi_{g_b g_b} \ .
\end{split}\end{equation}
These relations can be used to check or predict the  2-loop contribution to the gauge beta functions coming from the Yukawa interactions once the metric \eqref{eq:chis2l} is known.

\section{Local quantities at Fixed Points}
\label{Local}
\noindent
Assume that the theory described in \ref{eq:Lagrangian} admits an interacting fixed point. This phase of the theory is described by a CFT characterized by a well defined set of quantities. We loosely refer to this set as the \textit{conformal data} of the CFT and it includes the critical exponents as well as the quantities $a,c,a/c$. We refer the reader to Appendix~\ref{3loops} for definitions of central charges $a$ and $c$. Intuitively such quantities usually serve as a measure of degrees of freedom in the given CFT. In the present work, these are calculated using perturbation theory, so we rely on the assumption that the fixed point is not strongly coupled. This is usually under control in the Veneziano limit $N_c, N_f \rightarrow \infty$ such that the ratio  $\frac{N_f}{N_c}$ is finite. Moreover, such ratio needs to be close to the critical value for which the theory loses asymptotic freedom (which depends on the particular content of the theory). I.e we have $b_0 \propto \epsilon$ for some $\epsilon$ being a small positive parameter. This $\epsilon$ expansion is useful to determine the local quantities at the fixed point, and reorganize the perturbation theory series in the couplings.

\subsection{$\tilde{a}$-function at two loops}

When Weyl consistency conditions are satisfied we can integrate the gradient flow equation   \eqref{eq:consistencycondition} to determine  the lowest two orders of $\tilde{a}= \tilde{a}^{(0)} + \frac{\tilde{a}^{(1)}}{(4\pi)^2}+\frac{\tilde{a}^{(2)}}{(4\pi)^4}+\dots  $ the result is
\begin{equation}  \begin{split}
   \tilde{a}^{(0)}&  = \frac{1}{360 (4\pi)^2} \left( n_{\phi} + \frac{11}{2} n_{\psi} + 62 n_v \right) \ , \\
	\tilde{a}^{(1)} & =  -\frac{1}{2} \sum_a \chi_{g_a g_a} b_0^a g_a^2 \ ,
	\label{eq:atildeM:1} \\
			\tilde{a}^{(2)} & =  - \frac{1}{4}\sum_a \chi_{g_a g_a} g_a^2 \left[  \left( A_a b_0^a g_a^2+ \sum_b (b_1)^{ab}g_b^2\right)
		+ \sum_{IJ} (b_{y})_{ IJ}^a y^I y^J \right]
		+ 4\pi^2\sum_{I} \chi_{y_I y_I} \beta_{y}^I \ .\\
\end{split}\end{equation}
At fixed points $(g^{*},y_I^{*})$ this quantity becomes scheme-independent and physical and reduces to the so called \textit{a-function} (see \eqref{eq:atilde}) . This quantity partially characterizes the associated conformal field theory. The interacting fixed point requires  
 \begin{equation}\begin{split} 
 (c_2)^a_{IJ} (g_a^*)^2 y^{*J}& = -(c_1)_{IJKL}  y^{*J} y^{*K} y^{*L} \\
 (b_{y})_{IJ}^a y^{*I} y^{*J}  & = -(b_1)^{ab} (g_b^*)^2 - (4\pi)^2 b_0^a
 \label{zeroes}
\end{split}\end{equation} 
stemming from $\beta_g^a(g_a^{*},y_I^{*})=\beta_y^I(g_a^{*},y_I^{*})=0$ with the explicit expressions given in  \eqref{eq:betagaugeM},\eqref{eq:betaYukawaM}. We then have the general expression
\begin{equation}
\label{eq:AatFixed}
a^{*}= \tilde{a}^{*}= \afree - \frac{1}{4} \frac{1}{(4\pi)^2} \sum_a b_0^a \chi_{g_a g_a} g_a^{*2} \left(1 + \frac{A g_a^{*2}}{(4\pi)^2}\right) + \mathcal{O}(g_a^{*6},y_I^{*6}).
\end{equation}
The Yukawa couplings don't appear explicitly in the above expression to this order.  In the Veneziano limit the lowest order of $a$ behaves as $a^{(1)} \sim \epsilon^2$ since  $b_0^a \sim  g^{*2} \sim \epsilon$. The $a$ quantity has to satisfy the bound $a>0$ for any CFT.  Due to the fact that in perturbation theory the dominant term is the free one, positivity is ensured in large $N$ limit such that $\epsilon$ is arbitrary small. If the bound happens to be violated, it has to be interpreted as a failure of perturbation theory to that given order.

\subsection{$c$-function at two loops}
Adapting the two loop results given in \cite{Jack:1982sn, Osborn:2016bev}  to the generic theories envisioned here one derives for  $c=\frac{c^{(0)}}{(4\pi)^2}+\frac{c^{(1)}}{(4\pi)^4} + \dots $ the coefficients
\begin{eqnarray} \label{eq:c2loop}
	c^{(0)} & = & (4\pi)^2\cfree =  \frac{1}{20}(2 n_v +n_{\psi} + \frac{1}{6} n_{\phi} ) \ ,
	\\ \notag
	c^{(1)} &=&24 \left[ - \frac{2}{9} \sum_a g_a^2 d(\mathcal{G}_a) \left( C(\mathcal{G}_a) -\frac{7 }{16} \sum_i  T_{R^a_{\psi_i}} - \frac{1}{4} \sum_A  T_{R^a_{\phi_A}}\right) - \frac{1}{24} \sum_I (y^I)^2 Tr[T^I T^{I*}]\right] \ ,
\end{eqnarray}
where $T_{R^a_{\psi_i}} ,\,  T_{R^a_{\phi_A}}$ are the trace normalization for fermions and scalars  while $C(\mathcal{G}_a)$ is the Casimir of the adjoint representation. When this quantity is evaluated at the fixed point, it behaves as $c^{(1)} \sim \epsilon$. This order mismatch with $a$ was expected as $c$ does not satisfy a gradient flow equation in $d=4$.\\
Like the $a$ quantity, also $c$ is required to satisfy $c>0$ at a fixed point. Within perturbation theory however, the leading order is always positive definite so the violation of the bound is due to failure of the perturbative approach.

\subsection{ $a/c$ and collider bounds}
Having at our disposal both $a$ and $c$ one can discuss the quantity $\frac{a}{c}$. It has been shown in \cite{Hofman:2008ar,Parnachev:2008yt,Komargodski:2016gci,Hofman:2016awc} that the ratio of the  central charges $a$ and $c$ in $d=4$ satisfies the following inequality 
 \begin{equation} \label{eq:ColBounds}
 \frac{1}{3} \leq \frac{a}{c} \leq \frac{31}{18}\ , 
\end{equation}
which is known as {\it collider  bound}. Due to the fact that $a^{(1)} \sim \epsilon^2$ the next-to-leading order of $a/c$ takes contribution from $c^{(1)}$. 
To calculate the next correction we would need to know the $O(g^4)$ terms in $c$. Notice that because of  this order mismatch the ratio $\frac{a}{c}$ might become interesting also in perturbation theory for theories living at the edges of the collider bounds \eqref{eq:ColBounds}. For example free scalar field theories have $\frac{a}{c}=\frac{1}{3}$, this implies that for such theories the $\epsilon$ order coefficient has to satisfy $ a^{(0)}_{\epsilon} - c^{(0)}_{\epsilon} - c^{(1)}_{\epsilon}>0$. 


\subsection{Scaling Exponents} 

Is it always possible to linearize the RG flow in the proximity of a non-trivial fixed point and thus exaclty solve the flow equation
\begin{equation}
\beta^i (g^i) \sim \frac{\partial \beta^i}{\partial g^{j}} ( g^j - g^{j*}) + O(( g^j - g^{j*})^2) \implies g^i (\mu) = g^{i*} + \sum_k A_{k}^i c_k \left( \frac{\mu}{\Lambda} \right)^{\theta_{(i)}}
\label{powerscaling}
\end{equation}
where $\mu$ is the RG scale, $ c_k$ a is a coefficient depending on the initial conditions of the couplings, $A_{k}^i$ is the matrix diagonalizing $M_{j}^i = \frac{\partial \beta^i}{\partial g^{j}}$ and $\theta_{(i)}$ are the corresponding eigenvalues. These are called $critical \,\, exponents$ and from their signs one determines if the FP is UV/IR-attractive or mixed. It is worth notice that in the Veneziano limit, $M_{g}^g \sim \epsilon^2, M_{g}^I \sim \mathcal{O}(\epsilon^3)$, meaning that the mixing is not present at the lowest order and we will always have a critical exponent, say, $\theta_1$ such that $\theta_1 \sim M_g^g + \mathcal{O}(\epsilon^3) \sim \epsilon^2 + \mathcal{O}(\epsilon^3)$.

\section{Global properties of RG flows between fixed points}
\label{Global}

\subsection{ weak $a$-theorem}

The weak $a$-theorem states that, given a RG flow between a $CFT_{IR}$ and $CFT_{UV}$, the quantity $\Delta a = a_{UV} - a_{IR}$  is always positive \cite{Komargodski:2011vj}. This turns out to be a relevant constraint even in perturbation theory, as the zeroth order leading part cancels out. For example, for an arbitrary semi-simple gauge theory featuring either complete asymptotic freedom or infrared freedom the $\Delta a $ variation reads
\begin{equation}
\Delta a = \pm \frac{1}{4} \frac{1}{(4\pi)^2}  b_0^a \chi_{g g} g_a^{*2} \left(1 + \frac{A^b g_b^{*2}}{(4\pi)^2}\right) + \mathcal{O}(g_a^{*6},y_I^{*6}) \ .
\end{equation}
The plus (minus) applies when the theory is asymptotically free (infrared free). The case of the infrared free requires the ultraviolet theory to be asymptotically safe. \\
More generally, for single gauge coupling in Veneziano limit $|b_0| \ll 1$ we can derive a constraint from the leading order expression
\begin{equation} \label{eq:DaASAF}
\Delta a = - \frac{1}{4} \frac{1}{(4\pi)^2}  b_0 \chi_{g g} (g_{\text{UV}}^{*2}-g_{\text{IR}}^{*2} ) \geq 0 \; .
\end{equation}
Since $\chi_{gg}>0$, from the above inequality we derive the rather intuitive constraint that the gauge coupling has to increase (decrease) with the RG flow in asymptotically free(safe) theories. \\
A less intuitive constraint arises for theories featuring semi-simple gauge groups $ \otimes_{i=1,N} G_i$ for which we find
\begin{equation}\label{eq:DaSemisimple}
\Delta a = - \frac{1}{4} \frac{1}{(4\pi)^2} \sum_{i=1}^{N} b_{0}^a \chi_{g_a g_a} \Delta g_a^2 \geq 0 \; ,
\end{equation}
where $\Delta g_{a}^2= g_{\text{a UV}}^{*2}-g_{\text{a IR}}^{*2} $. From the above we obtain
\begin{eqnarray}
    \sum_i d(G_i)\, b_0^a\, \Delta g_a^2 \leq 0.
\end{eqnarray}
No other theorem is known to be valid for flows between two CFTs. For example it is known that $\Delta c$, in general, can be either positive or negative \cite{Cappelli:1990yc}. Let us conclude this subsection with a comment on the variation of the $\frac{a}{c}$ quantity. If one considers theories living at the edge of the collider bound, then within perturbation theory
\begin{equation}\label{eq:DelColQuant}
\Delta \left(\frac{a}{c} \right) \equiv \frac{a_{\text{UV}}}{c_{\text{UV}}}-\frac{a_{\text{IR}}}{c_{\text{IR}}}= - \frac{a_{\text{free}}}{c_{\text{free}}^2}\Delta c + \mathcal{O}(\epsilon^2) \ ,
\end{equation}
must be positive (negative) for the lower (upper) collider bound. This immediately translates in a bound for the $\Delta c$ sign. Of course, this is not expected to be valid beyond perturbation theory, while $\Delta a > 0$ was proven to hold even nonperturbatively \cite{Komargodski:2011vj}.

\subsection{Weakly relevant flows at strong coupling}
\label{sec:CritExp}

\begin{figure}[H]
\begin{center}
  \includegraphics[width=5 cm, scale=0.5]{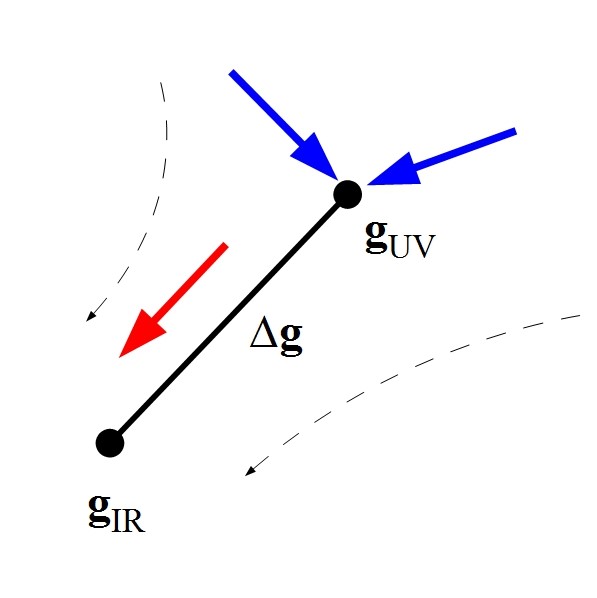}
  \caption{\small RG flow close to UV fixed point $\underline{g}_{UV} \equiv (g_{UV}^1, \dots g_{UV}^N)$. The thick black line represents the renormalized trajectory between two fixed points, which is parallel to the relevant direction (red arrow). Irrelevant directions correspond to blue arrows.  }
  \label{fig:Flow}
  \end{center}
\end{figure}

It is useful to discuss $\Delta a$ in the context of "conformal perturbation theory", which allows one to extend the perturbative analysis to potentially strongly coupled theories. The basic idea behind conformal perturbation theory is utilizing small deformation of CFT to study how does the behaviour close to fixed point depend on the its conformal data.  We will the assume existence of an interacting (not necessarily weakly coupled) CFT  in the UV and induce an RG flow by adding a slightly relevant coupling deformation. Utilizing this language we will derive the relation between $\Delta a$ of weakly relevant flows and critical exponents. \\
To set up the nomenclature we will consider a flow close to an arbitrary UV fixed point (denoted by $CFT_\text{UV}$) described by a set of $N$  couplings $g_\text{UV}^i$. We will deform the $\text{CFT}_\text{UV}$ slightly by $\Delta g^i$ with $|\Delta \underline{g}|\equiv (\Delta g^{i})^2\ll 1$  and assume, that within this regime there exists another fixed point $g_\text{IR}^i$ corresponding to $CFT_\text{IR}$. More concretely we will assume the existence of (diagonalized) beta functions in the vicinity of the fixed point (c.f. \eqref{eq:CPTbetas}) 
\begin{equation}
\label{eq:PertBetas}
\beta^i= \theta_{(i)} \Delta g^i + c_{jk}^i \Delta g^j\Delta g^k + \mathcal{O} (\Delta g^3)
\end{equation}
where $\theta_{(i)}$ correspond to critical exponents in the diagonalized basis of coupling space and $ c_{jk}^i$ are related to the OPE coefficients of associated nearly-marginal operators. In the following we will assume the existence of a nearby IR fixed point with $\Delta g^*$ such that $\beta^i(\Delta g^*)= 0 +  \mathcal{O} (\Delta g^3)$. A small $\Delta g^i$ solution exists if $\theta_{(i)}  \sim \epsilon \ll 1$ for  generic some small parameter $\epsilon$ (not necessarily equal to the Veneziano parameter) so that $\Delta g^* \sim \epsilon$ . \footnote{Note that if $c_{jk}^i$ is small (like it is the case in weakly coupled gauge theories), we need to expand $\beta^i$ to higher orders in orders to find a zero.}\\
Expanding the $\tilde{a}$- function close to $\underline{g}_\text{UV}$ we get
\begin{equation}
\label{eq:Aexpansion}
\tilde{a}_{\text{IR}}=\tilde{a}_{\text{UV}}+ \Delta g^{i} \partial_i \tilde{a}|_{\underline{g}_\text{UV}}+ \frac{1}{2} \Delta g^{i} \Delta g^{j} \partial_i \partial_j \tilde{a}|_{\underline{g}_\text{UV}} + \frac{1}{6} \Delta g^{i} \Delta g^{j} \Delta g^{k} \partial_i \partial_j \partial_k \tilde{a}|_{\underline{g}_\text{UV}} + \mathcal{O}(\Delta g^4)  \; .
\end{equation}
Next we will use the relation (see \eqref{eq:consistencycondition})
\begin{equation}
\label{eq:da}
\partial_i \tilde{a} \equiv \frac{\partial}{\partial g^i} \tilde{a}= \beta^j \chi_{i j}  \; ,
\end{equation}
where the metric $\chi_{ij}$ is positive close to fixed point \cite{Osborn:1989td}, \cite{Baume:2014rla} and we have assumed the one-form $w^i$ is exact close to a fixed point\footnote{This has been observed in all of the known examples. Most notably in perturbation theory close to a Gaussian fixed point in \cite{Jack:1990eb} and for supersymmetric theories in \cite{Auzzi:2015yia}. In two dimensions $w^i$ was proven to be exact \cite{Gomis:2015yaa}. }.  The equation \eqref{eq:da} also explains why we need to expand $a$ up to $(\Delta g)^3$. This is because the beta functions \eqref{eq:PertBetas} are $\mathcal{O}(\epsilon^2)$, so we expect their contribution to $\tilde{a}$ to be $\mathcal{O}(\epsilon^3)$.  \\
Using \eqref{eq:da} and the fact that beta functions have to vanish at the UV fixed point it is clear that the leading correction term ($\mathcal{O}(\Delta g)$ in \eqref{eq:Aexpansion}) vanishes and the we are left with
\begin{equation}
\label{eq:daConstrained}
\tilde{a}_{\text{IR}}=\tilde{a}_{\text{UV}}+ \frac{1}{2} \Delta g^{i} \Delta g^{j} \chi_{kj} \partial_i \beta^k |_{\underline{g}_\text{UV}}+ \frac{1}{6} \Delta g^{i} \Delta g^{j}  \Delta g^{k} \chi_{il} \partial_{j} \partial_{k} \beta^l |_{\underline{g}_\text{UV}} + \frac{1}{6}  \Delta g^{i} \Delta g^{j}  \Delta g^{k} \partial_{k} \chi_{i l} \partial_{j} \beta^l |_{\underline{g}_\text{UV}}     + \mathcal{O}(\Delta g^4)  \; .
\end{equation}
Note that the term proportional to $\partial \chi$ is $\mathcal{O}(\epsilon^4)$, hence by using \eqref{eq:PertBetas} we get
\begin{equation}
\label{eq:daConstrained1}
\tilde{a}_{\text{IR}}=\tilde{a}_{\text{UV}}+ \frac{1}{2} \Delta g^{i*} \Delta g^{j*} \chi_{ij} \theta_{(i)} + \frac{1}{3} \Delta g^{i*} \Delta g^{j*}  \Delta g^{k*} \chi_{il} c_{jk}^l     + \mathcal{O}(\epsilon^4)  \; .
\end{equation}
Now applying the fixed point condition $\beta^i(\Delta g^*)=0$ we get
\begin{equation}
\label{eq:daConstrained2}
\tilde{a}_{\text{IR}}=\tilde{a}_{\text{UV}}+ \frac{1}{6} \Delta g^{i*} \Delta g^{j*} \chi_{ij} \theta_{(i)}     + \mathcal{O}(\epsilon^4)  \; ,
\end{equation}
where we see that the OPE coefficients $c_{jk}^l$ dropped out at this order, so that the final result \eqref{eq:daConstrained2} only depends on the critical exponents. \\
Let us explore the RG flow close to UV fixed point (see Figure~\ref{fig:Flow}). In between the nearby fixed points, the renormalized trajectory can be described by a line joining the fixed points. In  known cases (E.g. \cite{Litim:2014uca}), this line is parallel to the direction corresponding to relevant eigenvector as indicated in Figure~\ref{fig:Flow}. If this is the case $\Delta \underline{g}$ is an eigenvector of the UV stability matrix $\partial_i \beta^k |_{\underline{g}_\text{UV}}$ and we clearly have 
\begin{equation}
\Delta g^{*i} \Delta g^{*j} \chi_{ij} \theta_{(i)}= \theta_{rel.}^{\text{UV}} (\Delta g^{* i})^2 \chi_{ii}  \; ,
\end{equation}
where $\theta_{rel.}^{\text{UV}} \equiv \theta_{(i)}$ is the critical exponent corresponding to the respective relevant direction on Figure~\ref{fig:Flow}. Therefore plugging this back into \eqref{eq:daConstrained2} we deduce
\begin{equation}
\label{eq:daConstrained}
\tilde{a}_{\text{IR}}=\tilde{a}_{\text{UV}}+ \frac{1}{6} \theta_{rel.}^{\text{UV}}(\Delta g^{* i})^2 \chi_{ii} + \mathcal{O}(\Delta g^3)  \; .
\end{equation}
Since $\theta_{rel.}^{\text{UV}}$ corresponds to a relevant direction it has to be negative so together with the positivity of $\chi_{ij}$ it  implies that to leading order the correction
\begin{equation}
\label{eq:DaRel}
\Delta a=\Delta \tilde{a}= \tilde{a}_{\text{UV}}-\tilde{a}_{\text{IR}}\approx -  \frac{1}{6} \theta_{rel.}^{\text{UV}}(\Delta g^{* i})^2 \chi_{ii}>0 \; ,
\end{equation}
consistently with the $a-$theorem. \\

\vskip .5cm 
The above result can be straight-forwardly extended to the case with multiple relevant couplings since we don't expect irrelevant directions to contribute to \eqref{eq:daConstrained2}.\\
We are now ready to provide the conformal data associated to distinct classes of asymptotically free or safe quantum field theories. 

\section{ The single Yukawa theory} \label{sec:1Yukawa}

We start with analysing the general model template featuring a simple gauge group and one Yukawa coupling. In the perturbation theory one can draw general conclusions on the phase diagram structure. At the 2-1-0 loop level two kinds of fixed points can arise: one in which both gauge and Yukawa couplings are non-zero (denoted as GY fixed point in the following) and a Banks-Zaks fixed point, where the gauge coupling is turned on while the Yukawa is zero (denoted as BZ fixed point). The control parameter $\epsilon$ in the Veneziano limit\footnote{This limit is strictly speaking applicable when considering SU($N_c$) gauge theories with matter in the fundamental representation. } is identified such that $b_0 \sim N_c \epsilon$, with $N_c$ the number of colours. These theories have the following general system of $\beta$-functions
\begin{equation}\label{eq:2loopBetas1c}\begin{split}
\beta_g &= -\frac{g^3}{(4\pi)^2} \Big[ b_0 + b_1 \frac{g^2}{(4\pi)^2} + b_y \frac{y^2}{(4\pi)^2} \Big] \\
\beta_y &=  \frac{y}{(4\pi)^2} \Big[ c_1 y^2 + c_2 g^2 \Big]
\end{split}\end{equation}
for which the following fixed points are present
\begin{equation} 
\left( \frac{g_{GY}^{2*}}{(4\pi)^2} = - \frac{b_0}{b_{1e}},\,  \frac{y_{GY}^{2*}}{(4\pi)^2} = \frac{c_2}{c_1}\frac{b_0}{b_{1e}}\right)  \qquad \left( \frac{g_{BZ}^{2*}}{(4\pi)^2} = - \frac{b_0}{b_1}, \frac{y_{BZ}^{2*}}{(4\pi)^2}  = 0\right)
\label{FP}
\end{equation}
where $b_{1e}=b_1  \Big(1-\frac{b_y}{b_1}\frac{c_2}{c_1}\Big)$. 
The above $\epsilon$-expansion of the fixed point couplings is reliable only up to $\mathcal{O}(\epsilon^2)$, where these higher orders are modified by higher loop corrections. These fixed points can be physical or not depending on the signs of the various beta function parameters.  \\
We will now calculate the conformal data for this general template to the leading 2-1-0 order . This corresponds to truncating every quantity to the first non trivial order in the $\epsilon$ expansion.

\subsubsection{scaling exponents}
\noindent
The scaling exponents at each fixed point are determined by diagonalising the rescaled flow matrix  $M_{ij} = \frac{1}{N_c} \frac{ \partial \beta_i}{ \partial g_j}$. These read
\begin{itemize}
\item BZ fixed point 
\begin{equation} \label{BZcriticalExponents}
\theta_1 =- 2\, \frac{b_0^2}{b_1} \sim \mathcal{O}(\epsilon^2),  \quad \theta_2 = - c_2\, \frac{b_0}{b_1} \sim \mathcal{O}( \epsilon )
\end{equation} 
The corresponding eigendirections are 
\begin{equation} 
v_1 = 
\left(\begin{array}{c}
1 \\
0
\end{array}\right) \quad
v_2 = 
\left(\begin{array}{c}
0 \\
1
\end{array}\right)
\end{equation}
And are thus parallel to the gauge-Yukawa coupling axis. Notice how the gauge coupling runs slower with respect to the Yukawa one, which therefore reaches asymptotic freedom much faster.
\item GY fixed point:  In general  $c_1 >0$  and  $c_2 <0$  \cite{Pica:2016krb} \\
\begin{equation}
\theta_1 = -2 \frac{b_0^2}{b_{1e} }\sim \mathcal{O}(\epsilon^2) \ , \quad \theta_2 =c_2\frac{b_0}{b_1} \sim \mathcal{O}(\epsilon)
\end{equation}
While the eigendirections are 
\begin{equation}
v_1 = 
\left(\begin{array}{c}
\frac{1}{\sqrt{1- \frac{c_2}{c_1}}}  -\frac{ b_{0}}{c_1 \sqrt{ (1-\frac{c_2}{c_1})^3}} + \mathcal{O}(\epsilon^2) \\
\frac{1}{\sqrt{1-\frac{c_1}{c_2}}} - \frac{c_1  b_{0}}{c_2^2 \sqrt{ (1+\frac{c_1}{c_2})^3}} + \mathcal{O}(\epsilon^2) \\
\end{array}\right) \quad
v_2 = 
\left(\begin{array}{c}
\frac{b_y b_0}{b_{1e} \sqrt{-c_1 c_2}} +\mathcal{O}(\epsilon^2) \\
1 + \frac{b_y^2 b_{0}^2}{2 c_1 c_2 b_{1e}^2} +\mathcal{O}(\epsilon^3)
\end{array}\right)
\end{equation}
Notice that  as $\epsilon \rightarrow 0 $ the flow between the GY fixed point and the Gaussian one becomes a straight line on the $v_1$ direction, forming an angle $\alpha$ with the $g$ axis such that $\tan ( \alpha ) = - \frac{c1}{c2}$. In this case a solution to the fixed flow equation is present. Moreover, since $\alpha \in [0,\pi/2]$ we see that if the GY fixed point is present then such a solution always exists, while the converse may not be true.\\
The eigencoupling along the direction of each eigenvector enjoys a power scaling close to the fixed point as in \eqref{powerscaling}, and the associated operator deformations then become either relevant or irrelevant depending on the sign of scaling exponents at the fixed point.
\end{itemize}

\subsubsection{Determining $a,c$ and the collider bound}
For the single gauge-Yukawa system \eqref{eq:2loopBetas1c} we can use the expressions \eqref{eq:atildeM:1},\eqref{eq:c2loop} to determine the  $a$,$c$ functions at fixed point. Notice that the $A$ coefficient has the expected $N_c$ dependence $A \sim N_c$. However, since the fixed point is known only to $\mathcal{O}(\epsilon)$ at two loop level, the $A$ term can be neglected since it only contributes to  $\mathcal{O}(\epsilon^3)$.  
We have 
\begin{itemize}
\item GY point
\begin{equation}
	a^* =\tilde{a}^*= \afree - \frac{1}{4}\chi_{gg} b_0 \frac{g^{*2}_{GY}}{(4\pi)^2} \,=\,  \afree - \frac{1}{8} \chi_{gg} \,\theta_1^{GY} + \mathcal{O}\left( \epsilon^3 \right) \ ,
\end{equation}
\begin{equation}
\label{eq:ASweakRelC}
c= c_{\text{free}} + \left(u\,- v\, \frac{c_2}{c_1} \right) \frac{b_0}{b_1}\,+  \mathcal{O}\left( \epsilon^2 \right)
\end{equation}
\begin{equation}
\frac{a}{c} = \frac{a_{F}}{c_{F}} \left[1 - \frac{1}{c_{F}}  \left(u\,- v\, \frac{c_2}{c_1} \right) \frac{b_0}{b_1} + \mathcal{O}(\epsilon^2) \right]
\end{equation}

\item BZ point
\begin{equation}
	a^* =\tilde{a}^* = \afree - \frac{1}{4}\chi_{gg} b_0 \frac{g^{*2}_{BZ}}{(4\pi)^2} \,=\,  \afree - \frac{1}{8} \chi_{gg} \,\theta_1^{BZ} + \mathcal{O}\left( \epsilon^3 \right) \ ,
\end{equation}
\begin{equation}
\label{eq:ASweakRelC}
c= c_{\text{free}} + u\, \frac{b_{0}}{b_1}\, \epsilon+  \mathcal{O}\left( \epsilon^2 \right) 
\end{equation}
\begin{equation}
\frac{a}{c} = \frac{a_{F}}{c_{F}} \left[1 - \frac{u\,}{c_{F}}\frac{b_0}{b_1} + \mathcal{O}(\epsilon^2) \right] \ .
\end{equation}

\end{itemize}

It is seen that for both of the above fixed points the two-loop contribution to the $a-$ function is proportional to the scaling exponent with the highest power in $\epsilon$
\begin{equation}
\Delta a = a_{FP} - a_{\text{free}} = - \frac{1}{8} \chi_{gg} \theta_{g}^{FP} +  \mathcal{O}\left( \epsilon^3 \right) \ .
\end{equation}
 The critical exponent in the above equation corresponds to the eigendirection pointing towards the Gaussian fixed point, which is coherent with our discussion in Section~\ref{sec:CritExp} for strongly coupled fixed points. This implies that for RG flows where one of the fixed points is Gaussian, we find again that $\Delta a$ is proportional to a scaling exponent.

\section{Related free and safe model templates}
\label{templates}

In the following we will calculate the local quantities for fixed point arising in different Gauge-Yukawa theories. We're interested in flows between an interacting fixed point and the Gaussian one. Depending on which point is the $CFT_{UV}$ these are either free or safe UV complete theories. We will consider these cases separately and provide examples for each one of them.

\subsection{Asymptotically Free Theories}
\subsubsection{{Vector-like SU(N) gauge-fermion theory}}

Consider an SU(N) gauge theory with vector-like fermions and its $\mathcal{N} = 1$ SYM extension, the field content is summarized in Table~\ref{table:QCD+SQCD}.
The supersymmetric extension of the model can be fitted into our gauge-Yukawa template introducing the following Yukawa interaction for each chiral field
\begin{equation}
\mathcal{L} = \left(\begin{array}{cc}
\psi_a & \lambda_A
\end{array}\right)
\left(\begin{array}{cc}
0 & \sqrt{2} g\,T^A_{ab} \\
\sqrt{2}g\, T^A_{ab} &0
\end{array}\right)
\left(\begin{array}{c}
\psi_a \\
\lambda_A 
\end{array}\right)\, \phi^b
+ h.c.
\end{equation}

 \begin{table}[H]
\[ \begin{array}{|c|c|c c|} \hline
{\rm Fields} &\left[ SU(N_c) \right] & SU_L(N_f) &SU_R(N_f) \\
 \hline 
\hline 
A_{\mu} & {\rm Adj} & 1 & 1   \\
 \psi &\Yfund &\Yfund & 1   \\
\tilde{\psi} & \overline{\Yfund} & 1 &  \Yfund   \\
\hline
\hline
\lambda & {\rm Adj} & 1 & 1   \\
 \phi &\Yfund &\Yfund & 1   \\
  \tilde{\phi} &\overline{\Yfund} & 1 & \Yfund    \\
 \hline
     \end{array} 
\]
\caption{Field content of the vector-like $SU(N)$ gauge theory. The lower table contains the superpartners of the $\mathcal{N}=1$ extension. }
\label{table:QCD+SQCD}
\end{table} 

These theories feature a Banks-Zaks fixed point arising at 2-loop level. The relevant beta function coefficients are known

\begin{equation}\begin{split}
&b_0^{\mathcal{N}=0} = \frac{2}{3} N_c \epsilon ; \quad b_1^{\mathcal{N}=0} = -\frac{25}{2} N_c^2+ \mathcal{O}(\epsilon); \quad \epsilon= \frac{\frac{11}{2}N_c-N_f}{N_c} >0 \; ,\\
&b_0^{\mathcal{N}=1} = N_c\epsilon  ; \quad b_1^{\mathcal{N}=1} = -6 N_c^2+ \mathcal{O}(\epsilon); \quad \epsilon= \frac{3N_c-N_f}{N_c} >0 \; .
\end{split}\end{equation}

Results are summarized in Table~\ref{table:BZn01}.\footnote{
The SUSY results of Table~\ref{table:BZn01} are readily confirmed by using the exact SUSY formulas \cite{Anselmi:1997ys}
\begin{equation}\begin{split}
c =\frac{1}{32} \frac{1}{(4 \pi)^2} \left[ 4 d(G) + d(r_i) \left(9 (R - 1)^3 - 5 (R - 1)\right)\right]\\
a= \frac{3}{32}\frac{1}{(4 \pi)^2} \left[2 d(G) +  d(r_i) \left(3 (R - 1)^3 - (R - 1)\right)\right]
\end{split}\end{equation}
with $R=\frac{2}{3} - \frac{\epsilon}{9}$ being the perturbative R-charge of squark field at the fixed point.}

 \begin{table}[H]
 \centering
\begin{tabular}{ |p{1.3cm}||p{1.5cm}|p{1cm}|p{1cm}|p{2.4cm}|p{1.5cm}|p{2cm}|  }\hline
  & $ \epsilon $ & $ \frac{N_c g^2}{ (4\pi)^2}$  & $\theta_g$ & $a \times \frac{(4 \pi)^2}{N_c^2}$ & $c \times \frac{(4 \pi)^2}{N_c^2}$ & $a/c$  \\
 \hline
 $  \mathcal{N} =0$  & $\frac{11}{2} - \frac{N_f}{N_c} $& $\frac{4\epsilon}{75}$  & $ \frac{16\epsilon^2}{225}$ & $\frac{49}{144} - \frac{11\epsilon}{360}- \frac{\epsilon^2}{225}$   & $\frac{288}{320}+\frac{19\epsilon}{80}$ & $ \frac{245}{468}- \frac{1933 \epsilon}{8112}$  \\
  $  \mathcal{N} =1$ & $ 3 - \frac{N_f}{N_c} $ & $\frac{\epsilon}{6}$  & $ \frac{\epsilon^2}{3}$ & $\frac{5}{16}-\frac{\epsilon}{24}- \frac{\epsilon^2}{48}$   & $\frac{3}{8} -\frac{\epsilon}{24}$ & $\frac{5}{6} - \frac{ \epsilon}{54}$  \\
  \hline
\end{tabular}\\
\caption{results for $\mathcal{N}=0,1$ gauge theories}
\label{table:BZn01} 
\end{table}
Additionally we have the expressions for global quantities
\begin{eqnarray}
\Delta a^{\mathcal{N} =0}&=& \frac{N_c^2}{(4 \pi)^2}\frac{1}{255} \epsilon^2 + \mathcal{O}(\epsilon^3) ;\\
\Delta a^{\mathcal{N} =1}&=& \frac{N_c^2}{(4 \pi)^2} \frac{1}{48} \epsilon^2 + \mathcal{O}(\epsilon^3) ;
\end{eqnarray}
The $\mathcal{N}=0$ agrees with the original result of \cite{Jack:1990eb}. We see at leading order the $a-$theorem doesn't provide any strong limits on $\epsilon$ so one might expect the higher orders will be more restrictive. However the recent $\mathcal{O}(\epsilon^4)$ evaluation of $\Delta a^{\mathcal{N} =0}$ in  \cite{Prochazka:2017pfa} reveals that to this order all the subleading coefficients remain to be positive providing no further perturbative bounds on $\epsilon$. 

\subsubsection{{Complete asymptotically free vector-like gauge theories with charged scalars}}
\label{section:scalars}

\begin{figure}[h] 
   \centering
   \includegraphics[width=3.in]{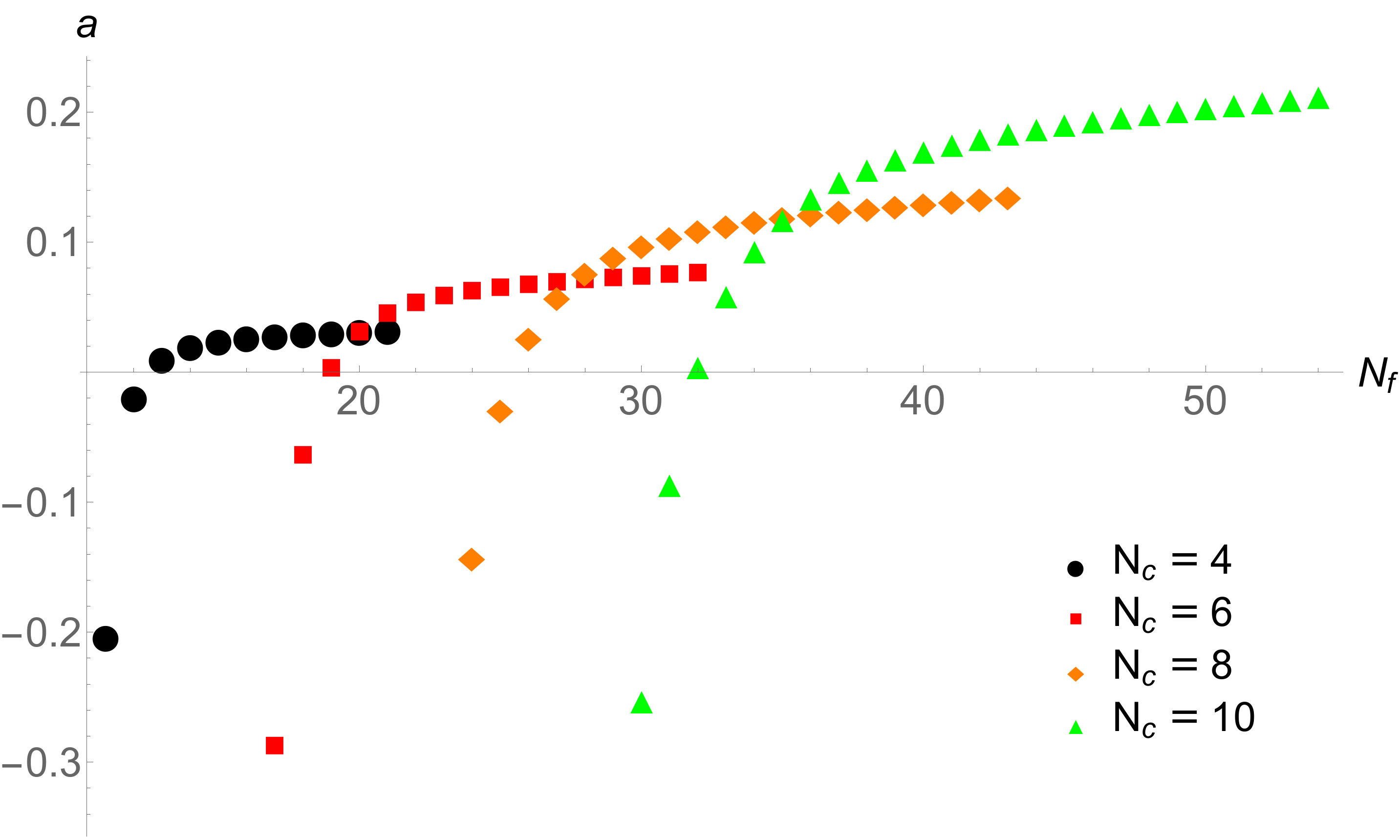} 
    \includegraphics[width=3.in]{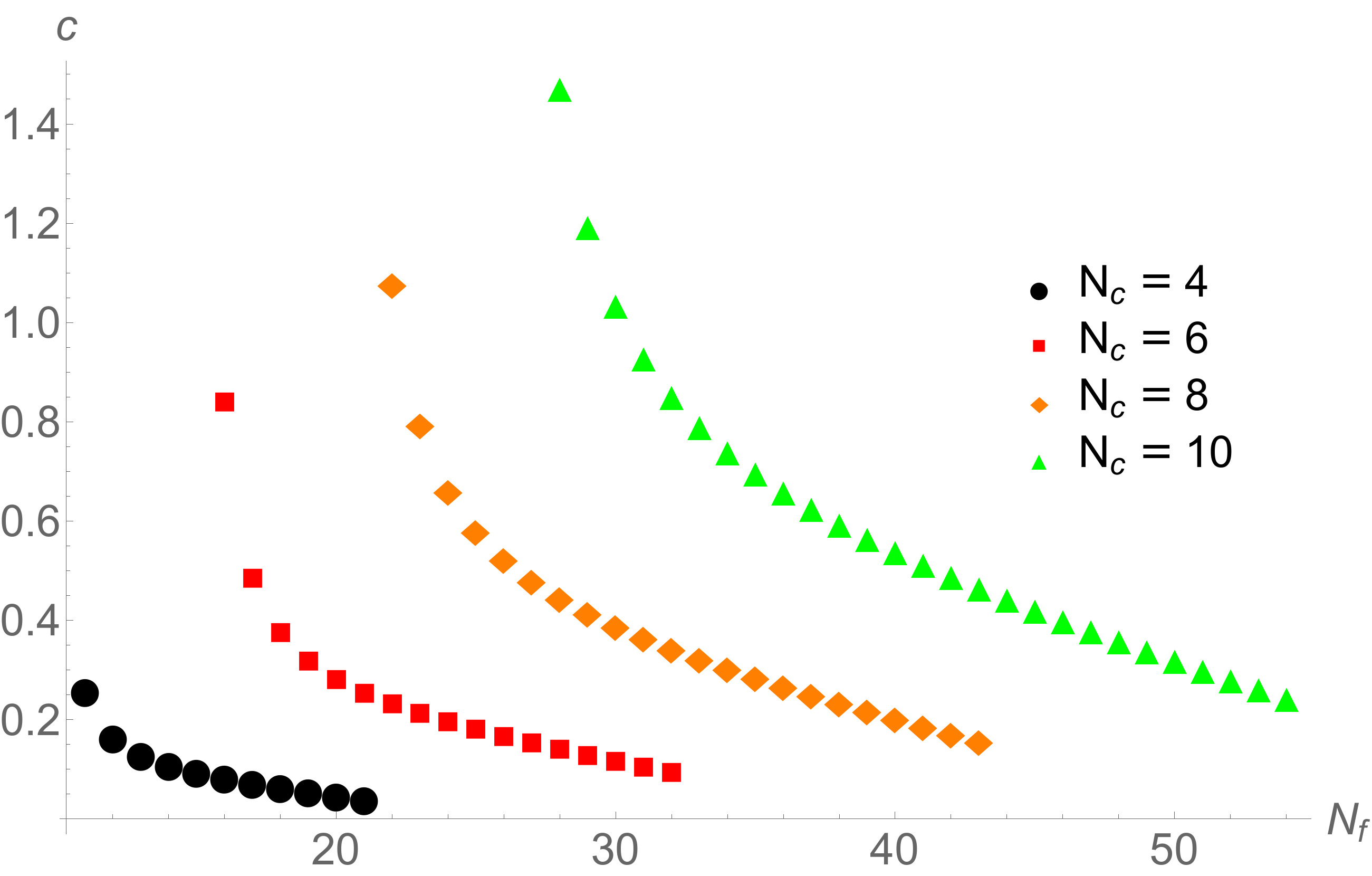} 
   \includegraphics[width=3.in]{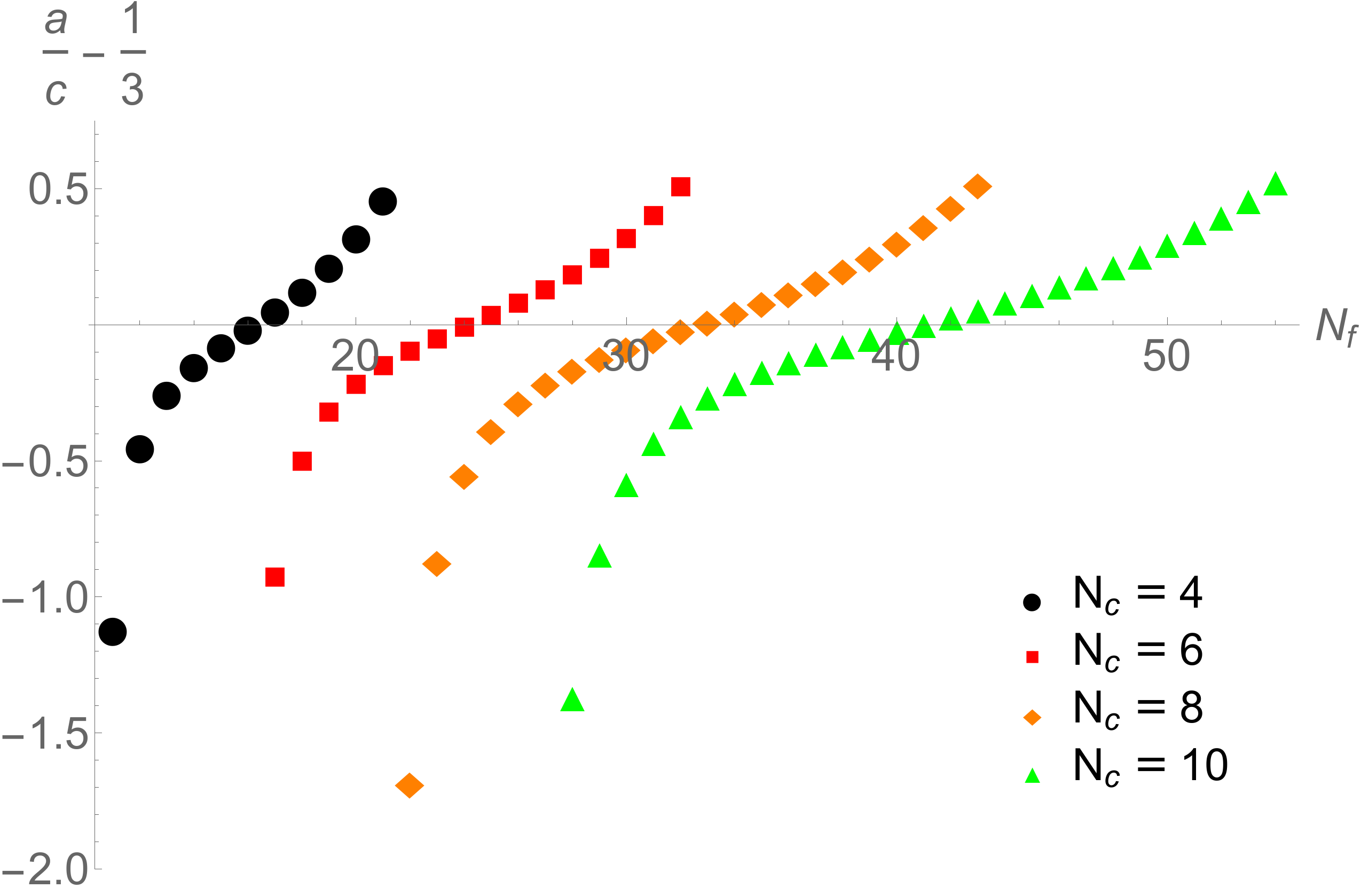} 
    \includegraphics[width=3.in]{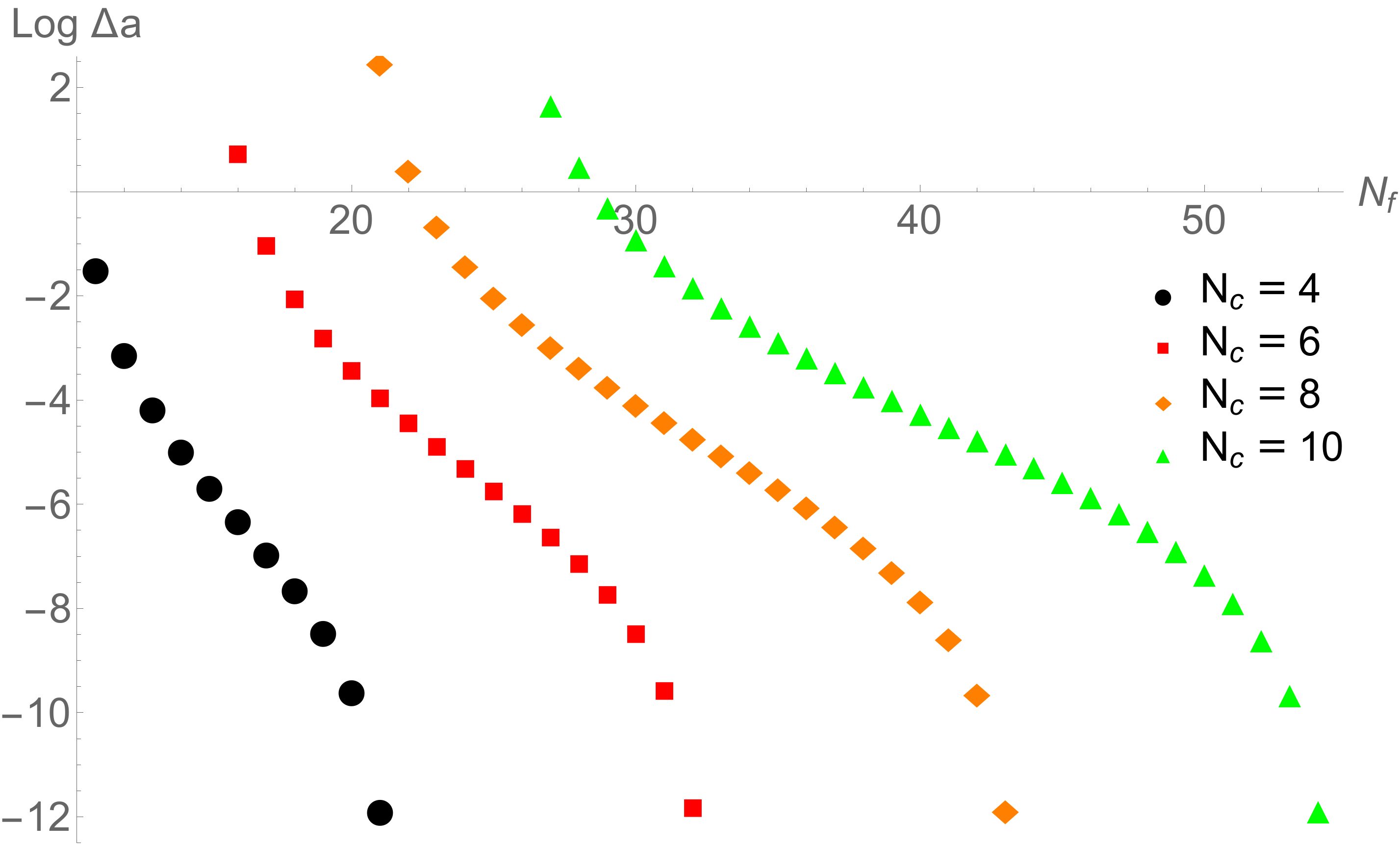} 
   \caption{$a$-function (upper left), $c$-function (upper right), collider bound $a/c$ (lower left) and $\Delta a$ (lower right) between the IR fixed point and the Gaussian. }
   \label{acac-scalars}
\end{figure}

Consider the scalar-gauge theory analyzed in \cite{Hansen:2017pwe} with matter content presented in Table~\ref{table:KasperContent}. Such model can be seen as the extension of the vector-like SU(N) gauge theory as well as the result of SUSY breaking of the $\mathcal{N}=1$ version with a scalar remnant.
 \begin{table}[H]
\[ \begin{array}{|c|c|c c c|} \hline
{\rm Fields} &\left[ SU(N_c) \right] & SU_L(N_f) &SU_R(N_f) & U(N_s)  \\
 \hline 
\hline 
 \psi &\Yfund &\Yfund & 1 & 1  \\
\tilde{\psi} & \overline{\Yfund} & 1 &  \Yfund  & 1 \\
\phi & \Yfund & 1 &  1  & \Yfund \\
\hline
     \end{array} 
\]
\caption{Field content of the model in \cite{Hansen:2017pwe} . }
\label{table:KasperContent}
\end{table} 
This model has no Yukawa couplings as they are forbidden by gauge invariance.
The scalar field features a self-interaction in the form of the usual single and double trace potentials
\begin{equation}
\mathcal{L} = -v\, \Tr[ \phi^{\dagger} \phi]^2 - u\, \Tr[(\phi^{\dagger} \phi)^2]
\end{equation}
It has been shown that this model features complete asymptotic freedom when an infrared fixed point is present. We analyze the flow between such point, when it exists, and the free UV one. At 2-1-1 loop level the fixed point splits into two  denoted as FP1, FP2 due to the presence of the scalar self-couplings and both of these are featuring a flow to the Gaussian fixed point. In Figure~\ref{acac-scalars} we plot the perturbative central charges of these fixed points for different vector-like flavours and colours. We focus on the minimal case realizing such fixed point, with number of complex scalars $N_s=2$. 
Notice that the central charges are evaluated at the two-loop level, so no distinction is present between FP1 and FP2 \cite{Antipin:2013pya}. We observe that the most sensitive quantity, as function of the number of flavours, is $a/c$, which fails to satisfy the lower bound $a/c > 1/3$ for sufficiently low number of flavours.
 $\Delta a$ is, however, always small and positive and spans several order of magnitude.\\
In Table~\ref{table:Kasper} we calculate positions of the fixed points and their critical exponents for the model in the large-$N_c , N_f$ limit of the model, where the central charges are identical to the ones on the first line of Table~\ref{table:BZn01}.

\begin{table}[H]
\centering
\begin{tabular}{ |p{0.8cm}|p{0.6cm}|p{2.5cm}|p{2.5cm}|p{4.5cm}|  }
 \hline
 &$ \frac{N_c g^{*2}}{ (4\pi)^2} $  & $ u^* N_f/(4\pi)^2$ & $ v^* N_f^2/(4\pi)^2$ & $ \theta_{(i)}$  \\
 \hline
 \hline
FP1 & $\frac{4 \epsilon }{75}$ & $-\frac{121}{300} (9-4 \sqrt{6}) \epsilon$ & $ \frac{11}{150} (3 - \sqrt{6}) \epsilon $ &
$  \frac{16}{225}\epsilon^2 \, ,
-\sqrt{\frac{2}{3}}\frac{8\epsilon}{25}\, ,
-\sqrt{\frac{2}{3}}\frac{8\epsilon}{25} $\\
FP2 & $\frac{4 \epsilon }{75}$ &  $-\frac{121}{300} (9+4 \sqrt{6}) \epsilon$ & $\frac{11}{150} (3 + \sqrt{6}) \epsilon $ & 
$  \frac{16}{225}\epsilon^2 \, ,
\sqrt{\frac{2}{3}}\frac{8\epsilon}{25}\, ,
\sqrt{\frac{2}{3}}\frac{8\epsilon}{25} $  \\
   \hline
   \end{tabular}
   \caption{Fixed points position and critical exponents in the Veneziano limit for the couplings of the theory.}
   \label{table:Kasper}
\end{table}

\subsubsection{{Complete asymptotically free chiral gauge-Yukawa theories}}

A further generalization of the previous models is obtained by adding chiral and vector-like fermions in higher dimensional representation of the gauge group. In particular we consider the models in \cite{Appelquist:2000qg,Molgaard:2016bqf}, namely the generalized Georgi-Glashow \cite{PhysRevLett.33.451} and Bars-Yankielowicz models \cite{BARS1981159} with the content of Table \ref{GG/BY} , which are by construction gauge anomaly free. We will work in the large $N_c$ limit tuning the constant $x=p/N_c$, so that these two theories are described by the same set of $\beta$-functions. 
At two-loop level this theory resembles  the template discussed in Section~\ref{sec:1Yukawa} where both BZ and GY fixed points are present.

 \begin{table}[H]
\[ \begin{array}{|c|c|c  c|} \hline
{\rm Fields} &\left[ SU(N_c) \right] & SU(N_c \mp 4 + p) &SU(p)  \\
 \hline 
\hline 
 \psi &\Yfund &1 & \Yfund   \\
\tilde{\psi} & \overline{\Yfund} & \overline{\Yfund} &  1   \\
A / S  & \Yasymm\, /\, \Ysymm & 1 &  1   \\
\hline
     \end{array} 
\]
\caption{Field content of the Georgi-Glashow / Bars-Yankielowicz models.}
\label{GG/BY}
\end{table} 
 
\begin{itemize}
\item \textbf{BZ fixed point.}\\
This type of fixed point arises for $\frac{3}{2} < x < \frac{9}{2}$ where at the lower limit it  becomes nonperturbative and at the upper one it merges with the Gaussian fixed point. We will thus expand around the perturbative edge of the $x$-window (also known as conformal window in the literature), namely write $x= \frac{9}{2} - \epsilon$ and arrive at the following fixed point
\begin{equation}
  \frac{N_c g^2_{BZ}}{(4\pi)^2} = \frac{2}{39} \epsilon + \mathcal{O}(\epsilon^2) \ , \qquad   \frac{N_c y^2_{BZ}}{(4\pi)^2} = 0 .
\end{equation}
At this fixed point we have the following set of eigendirection and critical exponents
\begin{equation}\begin{split}
&\theta_1 = \frac{8}{117} \epsilon^2 + \mathcal{O}(\epsilon^3)\quad \theta_2 = -\frac{4}{13}\epsilon + \mathcal{O}(\epsilon^2) \\
&v_1 = 
\left(\begin{array}{c}
1 \\
0 \\
\end{array}\right) \quad
v_2 = 
\left(\begin{array}{c}
\frac{11\epsilon}{26} \\
1-\frac{121\epsilon^2}{1352} \\
\end{array}\right) 
\end{split}\end{equation}
as well as the central charges values
\begin{equation}\begin{split}
a &=\frac{N_c^2}{(4 \pi)^2}\left(\frac{289}{720}-\frac{7 \epsilon }{120}-\frac{7 \epsilon^2 }{4680}\right)+\mathcal{O}(\epsilon^3)\\
c &=\frac{N_c^2}{(4 \pi)^2}\left(\frac{91}{160}+\frac{1193 \epsilon }{6240}\right)+\mathcal{O}(\epsilon^2)\\
a/c &=\frac{578}{819}-\frac{987670 \epsilon }{2906631}+\mathcal{O}(\epsilon^2)\\
\Delta a &= a_{FREE} - a_{BZ} = \frac{N_c^2}{(4 \pi)^2}\frac{\epsilon^2}{234} + \mathcal{O}(\epsilon^3)
\end{split}\end{equation}

\item \textbf{GY fixed point.}\\
This is present for $\frac{3}{8}(3+\sqrt{61}) < x < \frac{9}{2}$ and it behaves similarly to the BZ fixed point close to the upper and lower limits of the $x$ window. Working close to the upper edge of the conformal window $x= \frac{9}{2} - \epsilon$ we obtain
\begin{equation}
 \frac{N_c g^2_{GY}}{(4\pi)^2} = \frac{16}{15} \epsilon + \mathcal{O}(\epsilon^2)\ , \qquad   \frac{N_c y^2_{BZ}}{(4\pi)^2} = \frac{8}{15} \epsilon + \mathcal{O}(\epsilon^2) 
\end{equation}
at which we have the following set of eigendirection and critical exponents
\begin{equation}\begin{split}
&\theta_1 = \frac{64}{45} \epsilon^2 + \mathcal{O}(\epsilon^3)\quad \theta_2 = \frac{32}{5}\epsilon + \mathcal{O}(\epsilon^2) \\
&v_1 = 
\left(\begin{array}{c}
\sqrt{\frac{4}{5}}-\frac{7\epsilon}{75\sqrt{5}}  \\
\sqrt{\frac{1}{5}}+\frac{14\epsilon}{45\sqrt{5}} \\
\end{array}\right) \quad
v_2 = 
\left(\begin{array}{c}
-\frac{44}{5} \epsilon \\
1-\frac{968}{25} \epsilon^2 \\
\end{array}\right) 
\end{split}\end{equation}
The central charges are now
\begin{equation}\begin{split}
a &=\frac{N_c^2}{(4 \pi)^2}\left(\frac{289}{720}-\frac{7 \epsilon }{120}-\frac{31 \epsilon^2 }{360}\right)+O\left(\epsilon^3\right)\\
c &=\frac{N_c^2}{(4 \pi)^2}\left(\frac{91}{160}+\frac{761 \epsilon }{120}\right)+\mathcal{O}(\epsilon^2)\\
a/c &=\frac{579}{819}-\frac{1782364  \epsilon }{223587}+\mathcal{O}(\epsilon^2)\\
\Delta a &= a_{FREE} - a_{GY} = \frac{N_c^2}{(4 \pi)^2}\frac{4\epsilon^2}{45} + \mathcal{O}(\epsilon^3)
\end{split}\end{equation}

One can notice that a flow between the two non trivial fixed points is present, in which the BZ fixed point can be viewed as the UV completion of the GY one. This is supported by the positivity of $\Delta a$ between these two points
\begin{equation}
\Delta a = a_{BZ} - a_{GY} = \frac{N_c^2}{(4 \pi)^2} \frac{11 \epsilon^2}{130}.
\end{equation}

 \end{itemize}

\subsection{Safe models}

The discovery of asymptotic safety in four dimensions \cite{Litim:2014uca} has triggered much interest. It is therefore timely to investigate the associated conformal data. 

\subsubsection{{SU(N) with $N_f$ fundamental flavours and (gauged) scalars }}
\label{sec:LSmodel}

We start with the original theory that we will refer to, in the following, as LS theory  \cite{Litim:2014uca} that features the following field content   
  \begin{table}[H]
\[ \begin{array}{|c|c|c  c|} \hline
{\rm } &\left[ SU(N_c) \right] & SU(N_f)_L & SU(N_f)_R  \\
 \hline 
\hline 
 \psi &\Yfund &\overline{\Yfund} & 1   \\
\tilde{\psi} & \overline{\Yfund} & 1 &  \Yfund   \\
\phi  & 1 & \Yfund &  \overline{\Yfund}   \\
\hline
     \end{array} 
\]
\caption{Field content of the LS model.}
\label{LSmodel}
\end{table} 
and the Lagrangian
 \begin{equation}\begin{split}
& \mathcal{L}_{Y} = y\, \psi \phi \tilde{\psi}  + h.c.\\
& \mathcal{L}_S = - u \Tr[ (\phi^{\dagger} \phi)^2] - v (\Tr[ \phi^{\dagger} \phi ])^2 
\end{split} \end{equation}
As before at 2-1-0 order we will only focus on Yukawa coupling, keeping $N_c,N_f$ large. This time we will consider $0<\frac{N_f-\frac{11}{2} N_c}{N_c}= \epsilon \ll 1$, slightly above the asymptotic freedom bound. Such theory possesses an UV fixed point \cite{Litim:2014uca}. In the Veneziano limit the coefficients of \eqref{eq:2loopBetas1c} read
\begin{eqnarray} \notag
b_0 =& - \frac{2}{3} \epsilon N_c ; \quad &b_1 = -\left(\frac{25}{2} - \frac{13}{3} \epsilon \right) N_c^2 ; \quad b_y = \frac{121}{4}  N_c^2 + \mathcal{O}(\epsilon);   \\
c_1 =&  \frac{13}{2}N_c + \mathcal{O}(\epsilon)  ; \quad &c_2 = -3 N_c 
\quad \ .
\end{eqnarray}
Therefore we have $b_{1e}= \frac{19}{13} N_c^2$, which leads to the following UV fixed point \cite{Litim:2014uca}
\begin{equation}
\left( \frac{N_c g^{*2}}{(4\pi)^2} = \frac{26}{57} \epsilon, \,  \frac{ N_cy^{*2}}{(4\pi)^2} = \frac{12}{57}\epsilon \right) \quad \ .
\label{FPLS}
\end{equation}
The critical exponents yield
\begin{equation}
\label{eq:LScritExp}
\theta_1= - 2 \frac{b_0^2}{b_{1e}}= -\frac{104}{171}\epsilon^2 ; \quad \theta_2=  2 c_2 \frac{b_0}{b_{1e}}= \frac{52}{19} \epsilon ;
\end{equation}
corresponding to the eigendirections 
\begin{equation}
v_1= \left(\begin{array}{c}
\sqrt{\frac{13}{19}} \\
\sqrt{\frac{6}{19}} 
\end{array}\right) ; \quad v_1= \left(\begin{array}{c}
0 \\
1 
\end{array}\right) \; .
\end{equation}
The $a$ function at this fixed point is given by
\begin{equation}
a_{\text{LS}}= \afree - \frac{1}{4} \frac{\chi_{gg}}{(4\pi)^4}  b_0 g^{*2}= \afree + \frac{13N_c^2}{342} \frac{1}{(4\pi)^2} \epsilon^2 =\frac{1}{(4\pi)^2}\frac{N_c^2}{120}\left(61+ 11 \epsilon + \frac{298 \epsilon^2}{57} \right)\quad \ .
\end{equation} 
Next we will proceed to calculate $c_{\text{LS}}$  
\begin{equation}
c_{\text{LS}}= \cfree + \frac{31N_c^2}{68} \frac{1}{(4\pi)^2} \epsilon= \frac{N_c^2}{(4\pi)^2}\frac{1}{120}\left(\frac{211}{2}+ \frac{2 \epsilon}{17}+ \mathcal{O}(\epsilon^2) \right)  \quad \ .
\end{equation} 
Note we would need to know the $\mathcal{O}(g^4, y^4)$ contribution to $c$, in order to determine the $\mathcal{O}(\epsilon^2)$ correction of the $a/c$ quantity that to order $\epsilon$ reads:
\begin{equation}
\frac{a_{LS}}{c_{LS}} = \frac{122}{211}+ \frac{78426  \epsilon}{756857} + \mathcal{O}(\epsilon^2) \ .
\end{equation}
Notice that the collider bound is well satisfied as long as $\epsilon \lesssim 1$. Using the general result obtained in Section~\ref{sec:beyond2loops} it is possible to obtain the 3-loop expression for $\Delta a$ between the UV safe fixed point and the Gaussian one in the IR:
\begin{equation}
\Delta a = \frac{N_c^2}{(4 \pi)^2}\left[\frac{13}{342} \epsilon^2 + \left(\frac{65201-11132 \sqrt{23}}{246924}\right)  \epsilon^3 \right] + \mathcal{O}(\epsilon^4) \ .
\end{equation}
Even at finite $N_c$ and $N_f$ asymptotic safety abides the local and global constraints as long as the $\epsilon$ parameter is controllably small.

Recently, this model has been extended \cite{Pelaggi:2017wzr} to accommodate a gauged Higgs-like scalar (in fundamental representation) and $2N_f$ singlet fermions $N^i, N'^i$ as summarized in Table \ref{Strumia}.
 \begin{table}[H]
\[ \begin{array}{|c|c|c  c|} \hline
{\rm Fields} &\left[ SU(N_c) \right] & SU(N_f)_L & SU(N_f)_R  \\
 \hline 
\hline 
 \psi &\Yfund &\overline{\Yfund} & 1   \\
\tilde{\psi} & \overline{\Yfund} & 1 &  \Yfund   \\
\phi  & 1 & \Yfund &  \overline{\Yfund}   \\
\hline
\hline
 H &\Yfund & 1 & 1   \\
N & 1 & 1 &  \overline{\Yfund}   \\
N'  & 1 & \Yfund & 1  \\
\hline
     \end{array} 
\]
\caption{Field content of the model of \cite{Pelaggi:2017wzr} .}
\label{Strumia}
\end{table} 
This theory has some extra Yukawa and scalar couplings 
\begin{equation}\begin{split}
& \mathcal{L}_{Y} = y\, \psi \phi \tilde{\psi} + y' N H^{\dagger} N' + \tilde{y} H \tilde{\psi} N + \tilde{y}' H^{\dagger} \psi N' + h.c.\\
& \mathcal{L}_S = - \lambda_{S1} \Tr[ \phi^{\dagger} \phi ]^2 - \lambda_{S2} \Tr[ (\phi^{\dagger} \phi)^2 ] - \lambda_H (H^{\dagger} H)^2 - \lambda_{HS} (H^{\dagger}H) \Tr[ \phi^{\dagger} \phi]
\end{split} \end{equation}
Note that beta functions of these 3 new Yukawa couplings decouple in the Veneziano limit.
The fixed point found in \cite{Pelaggi:2017wzr} appears at $y^{'*}= \tilde{y}^{'*}=0$ and $\frac{N_c\tilde{y}^{*2}}{(4\pi)^2}= \frac{\epsilon}{26}$. Since at the 2-1-0 level $\beta_{g,y}$ doesn't depend on $\tilde{y}$, the model enjoys the LS critical exponents \eqref{eq:LScritExp} with the third one being
\begin{equation}
\theta_3= \frac{\partial}{\partial \tilde{y}} \beta_{\tilde{y}}|_{g^*,y^*,\tilde{y}^*}= \frac{6}{13} \epsilon \ ,
\end{equation}
which corresponds to an extra irrelevant direction in the coupling space. \\
Clearly the $a-$function of this model is identical to the LS one since both models have the same $b_0$ (c.f. \eqref{eq:AatFixed}). Similarly the $c-$ function of this model is identical to the LS one. This is due to the fact that the extra contribution of $\tilde{y}$ in \eqref{eq:c2loop} is proportional to $\Tr (T_{\tilde{y}} T_{\tilde{y}}^*) \propto N_c N_f$  which is suppressed in the Veneziano limit compared to the $g , y$ contribution proportional to $\Tr (T_y T_y^*) \propto N_c N_f^2$ so we can neglect it.

\subsubsection{{Complete asymptotically safe chiral models}}
The UV dynamics of Georgi-Glashow (GG) models that include also singlet as well as charged scalar fields was investigated in \cite{Molgaard:2016bqf}. 
 \begin{table}[H]
\[ \begin{array}{|c|c|c  c|} \hline
{\rm Fields} &\left[ SU(N_c) \right] & SU(N_c \mp 4 + p) &SU(p)  \\
 \hline 
\hline 
 \psi &\Yfund &1 & \Yfund   \\
\tilde{\psi} & \overline{\Yfund} & \overline{\Yfund} &  1   \\
A   & \Yasymm\,  & 1 &  1   \\
\hline
\hline 
 M & 1  & \Yfund & \overline{\Yfund}   \\
H & \overline{\Yfund} & 1 &  1   \\
\hline
     \end{array} 
\]
\caption{Field content of the Georgi-Glashow models extended with singlet and charged scalars.}
\label{GG+scalar}
\end{table} 

The field content is summarized in Table~\ref{GG+scalar} and the interactions between chiral fermions and scalars are described via the following Lagrangian terms
\begin{equation}\begin{split}
&\mathcal{L}_H = y_H f_a\, \bar{\psi}_a A\, H + h.c. \\
&\mathcal{L}_M =y_M [ \delta_{ab} -f_a f_b ]\,\bar{\psi}_a M_{bc}\psi_c + y_1 f_a f_b\, \bar{\psi}_a M_{bc}\psi_c + h.c.
\end{split}\end{equation}
Where $f_a$ is a vector in flavour space. The Higgs-like scalar breaks the flavour symmetry with the Yukawa term $y_H$. In the following we choose to have just one flavour interacting with the $H$ field, so $f_a = \delta_{a,1} $. The distinction between $y_M,y_1$ is convenient as loop corrections will differentiate between the flavour interacting with $H$ with the others.\\
It is possible to show that the Bars-Yankielowicz (BY)\footnote{The difference w.r.t. to the Georgi-Glashow theories is that the Weyl fermion transforming according to the two-index antisymmetric tensor under the gauge group is replaced by a symmetric one.} version of the theory cannot lead to complete asymptotic safety for any $N_c$. Within the GG,  the fully interacting FP of this theory at 2-1-0 loop level is fully IR attractive in the large $N_c$ limit. However there are some candidates of finite $N_c$ theories for which complete asymptotic safety can potentially emerge. We now determine the conformal data for the three  candidate fixed points found in the original work, these are shown in Table~\ref{table:GG-CAS}.\\
We find that all these UV fixed points, at least in some of the couplings,  are clearly outside the perturbative regime given that $a/c$ and $\Delta a$ constraints are not respected. 
\begin{table}[H]
\begin{center}
\begin{tabular}{ |p{1.3cm}||p{2.5cm}|p{2.5cm}|p{2.5cm}|  }
 \hline
      & $N_c =5,\, p=26$ & $N_c =6,\, p=30$  & $N_c =8,\, p=39$ \\
 \hline
$\alpha_g^*$  &1.41 & 0.0325  &  0.0481\\
\hline
$\alpha_H^*$  & 6.12  & 0.151 &0.241 \\
\hline
$\alpha_M^*$  & 0.652 & 0.0155& 0.0233 \\
\hline
$\alpha_1^*$  & 0.312 & 0.00652 & 0.00801  \\
\hline
$\theta_{UV}$  & -0.0428 & -0.00585 & -0.00602\\
\hline
$a\times(4 \pi)^2$  &  -1311  & 14.7 & 21.6 \\
\hline
$c\times(4 \pi)^2$  & 710  & 47.5 & 126 \\
\hline
$a/c$  & -1.84  & 0.296 & 0.171\\
 \hline
 $\Delta a$  & -1321 & -0.537 & -4.27 \\
 \hline
\end{tabular}
\caption{couplings, critical exponents and central charges for fixed points that can realize complete asymptotic safety (CAS) .}
\label{table:GG-CAS}
\end{center}
\end{table}

\subsection{Flows between interacting fixed points}

Here we would like to consider models possessing interacting fixed points in both  IR and UV. In the following we will investigate the $a-$theorem constraints to further characterise such flows.

\subsubsection{{BZ-GY flow in the completely asymptotically free regime}}

Let us now turn to a class of theories with 
\begin{eqnarray} \label{eq:ComplAFcond}
b_0>0 , \quad b_1<0, \quad b_{1e}<0 .    
\end{eqnarray}
The main features of these models were discussed in Section~\ref{sec:1Yukawa}. We also refer the reader to \cite{Pica:2016krb} for a more detailed discussion.  A concrete example can be realized by coupling the LS model (c.f. Section~\ref{sec:LSmodel}) to some additional fermions in the adjoint representation. Clearly, if the conditions \eqref{eq:ComplAFcond} are satisfied, both GY and BZ fixed points (c.f. \eqref{FP}) can coexist. Furthermore if $c_2<0$, the BZ fixed point acquires a relevant direction corresponding to the Yukawa coupling (see \eqref{BZcriticalExponents}). It is therefore reasonable to expect, that there's an RG flow between BZ and GY points. Indeed, using \eqref{eq:DaASAF} we find that for a generic gauge theory with group $G$ (recall that $\chi_{gg}= \frac{1}{2} \frac{1}{4 \pi^2} d(G)$) we have
 \begin{equation} \label{eq:delaAF}
     \Delta a= a_{\text{BZ}}- a_{\text{GY}}= - \frac{1}{8} d(G) \frac{b_0^2}{b_1^2} \frac{b_y\frac{c_2}{c_1}}{\left(1- \frac{b_y}{b_1}\frac{c_2}{c_1} \right)} \; ,
 \end{equation}
 which is positive since $b_{1e}<0$ implies
 \begin{eqnarray}
     b_y \frac{c_2}{c_1}<b_1<0 \; .
 \end{eqnarray}
 More concretely we can take an extension of the model described in Section~\ref{sec:LSmodel} with an extra gluino-like adjoint fermion (see Table \ref{GG/AFInt}).
 \begin{table}[H]
\[ \begin{array}{|c|c|c  c|} \hline
{\rm Fields} &\left[ SU(N_c) \right] & SU(N_f)_L & SU(N_f)_R  \\
 \hline 
\hline 
 \psi &\Yfund &\overline{\Yfund} & 1   \\
\tilde{\psi} & \overline{\Yfund} & 1 &  \Yfund   \\
\phi  & 1 & \Yfund &  \overline{\Yfund}   \\
\hline
\hline
 \lambda &{\rm Adj} & 1 & 1   \\
\hline
     \end{array} 
\]
\caption{Field content of the LS model with an additional adjoint Weyl fermion.}
\label{GG/AFInt}
\end{table}  
 The relevant beta function coefficients in the Veneziano limit read
\begin{eqnarray} \notag
b_0 =&  \frac{2}{3} \epsilon N_c ; \quad &b_1 = -\frac{27}{2} N_c^2 ; \quad b_y = \frac{81}{4} ; \quad \epsilon= \frac{\frac{9}{2}N_c-N_f}{N_c} >0 ;\\
c_1 =&  \frac{11}{2}N_c + \mathcal{O}(\epsilon)  ; \quad &c_2 = -3 N_c 
\quad \ .
\end{eqnarray}
For this model $b_{1e}=-\frac{27}{11} N_c^2$ and hence it satisfies the complete asymptotic safety criterion with two fixed points
\begin{equation}
 \frac{N_c g^{2}_{GY}}{(4\pi)^2} = \frac{22}{81} \, \epsilon + \mathcal{O}(\epsilon^2) ,\quad  \frac{N_c g^{2}_{BZ}}{(4\pi)^2} = \frac{4}{81}\, \epsilon + \mathcal{O}(\epsilon^2)
\label{FPaf}
\end{equation}
The flow between these two fixed point (where BZ plays the role of UV fixed point) satisfies the $a-$theorem \eqref{eq:delaAF}
\begin{eqnarray}
    \Delta a= \frac{1}{(4 \pi^2)}\frac{\epsilon^2 N_c^2}{54} ,
\end{eqnarray}
which is positive as expected from the above discussion.

\subsubsection{{Large $N_f$ safety with the Higgs: }}

Recently an interesting class of large $N_f$ models with strongly-coupled UV fixed point has been discussed in the literature \cite{Mann:2017wzh,Abel:2017rwl,Pelaggi:2017abg}. These models extend previous work \cite{Holdom:2010qs,Pica:2010xq}  by including the Higgs and therefore  provide a realistic framework for asymptotically safe extensions of the standard model. Further insights on the nature and consistence of these fixed points were investigated in \cite{Antipin:2017ebo}.\\
 Here we  consider a model with large $N_f$ vector-like fermions and a Higgs-like scalar transforming according to the fundamental representation of $SU(N_c)$. The field content of this theory is summarized in the following table.
  \begin{table}[H]
\[ \begin{array}{|c|c|c  c|} \hline
{\rm Fields} &\left[ SU(N_c) \right] & SU(N_f)_L & SU(N_f)_R  \\
 \hline 
\hline 
 \psi &\Yfund &\overline{\Yfund} & 1   \\
\tilde{\psi} & \overline{\Yfund} & 1 &  \Yfund   \\
\phi  & 1 & \Yfund &  \overline{\Yfund}   \\
\hline
\hline
 H &\Yfund & 1 & 1   \\
\hline
     \end{array} 
\]
\caption{Field content of the model.}
\label{LargeNf}
\end{table}  
  Expanding to leading order in  $\frac{1}{N_f}$, the gauge beta function of this theory has a zero at
\begin{eqnarray}
    A^*= \frac{N_f g^{*2}}{(4 \pi)^2}= 3 + (\dots)e^{-\frac{Nf}{Nc}} \; .
\end{eqnarray}
The presence of the Higgs self-coupling $\lambda_H$ does not disturb this fixed point at this order in $\frac{1}{N_f}$ \cite{Abel:2017rwl} (up to exponentially suppressed contributions). At one loop the gauge coupling appears without powers of $N_f$ in the quartic beta function, which is therefore under perturbative control. To make the calculation more manageable we can also take the intermediate large $N_c$ limit provided \cite{Antipin:2017ebo}
\begin{equation}
    \frac{1}{10} > \frac{N_c}{N_f} \gg \frac{1}{N_f} \; .
\end{equation}
Thus given we keep $\frac{N_c}{N_f}$ small enough, this regime is attainable. To keep track of $\frac{1}{N_f}$ in the scalar sector we introduce the coupling
$u_H= \frac{\lambda_H N_f}{(4\pi)^2}$ with the following beta function (close to the above UV fixed point $A \to A^*$) 
\begin{eqnarray}
    \beta_{u_H}= \frac{1}{N_f} \left(4 N_c u_H^2- 6N_c u_H A^* + \frac{3}{4} N_c A^{*2}   + \mathcal{O}\left(\frac{1}{N_f^2}, \frac{1}{N_c}\right) \right)\; .
\end{eqnarray}
Thus we see that this beta function remains small if we keep $N_f$ large enough. Neglecting the subleading contributions we have
\begin{eqnarray}
    \beta_{u_H}= \frac{4 N_c}{N_f} (u_H-u_+)(u_H- u_-) \; ,
\end{eqnarray}
where
\begin{eqnarray}\label{eq:u=-}
    u_{\pm}= \left( 3 \pm \sqrt{6} \right) \frac{A^*}{4 N_f} \; .
\end{eqnarray}
This allows for two fixed points with stable Higgs potential since $u_\pm >0$. Furthermore, $u_-$ is fully UV-attractive since the corresponding critical exponent satisfies \footnote{Note that the critical exponent corresponding to gauge coupling is always negative irrespectively of the presence of scalar.}
\begin{eqnarray} \label{eq:theta-}
    \theta_{u_-}= \frac{\partial}{\partial u_H} \beta_{u_H}|_{u_-}= - 4 \frac{N_c}{N_f}(u_+ - u_-)= - 2 \sqrt{6} \frac{N_c}{N_f} A^*= - 6 \sqrt{6} \frac{N_c}{N_f} < 0 \; .
\end{eqnarray}
By deforming the scalar coupling away from the fixed point $\delta u_H= (u_H-u_-)>0$ and keeping the gauge coupling fixed, we expect the theory to flow to a new (also strongly coupled) IR fixed point at $u_H=u_+$. We are now ready to compute $\Delta a$ for this flow using \eqref{eq:DaRel} and the leading large $N_f$ behaviour of the metric $\chi$
\begin{equation} \label{eq:chiuu}
    \chi_{u_H u_H}= \frac{1}{24 } \left(\frac{N_c}{N_f}\right)^2\frac{1}{(4\pi)^2} + \mathcal{O}\left(\frac{1}{N_f^3}\right) \; .
\end{equation}
Plugging the quantities \eqref{eq:u=-}, \eqref{eq:theta-}, \eqref{eq:chiuu} directly into \eqref{eq:DaRel} we obtain the final result
\begin{eqnarray}
    \Delta a = -\frac{1}{6}\chi_{u_H u_H} \theta_{u_-} (u_+ - u_-)^2+ \mathcal{O}\left(\frac{1}{N_f^4}\right)= \frac{1}{(4\pi)^2} \left(\frac{N_c}{N_f}\right)^3 \frac{9 \sqrt{6}}{16}+ \mathcal{O}\left(\frac{1}{N_f^4}\right) \; .
\end{eqnarray}
Note that the smallness of $\Delta a$ is controlled by $\frac{N_c}{N_f}$, which is kept small.

\section{Conclusions}
\label{conclusions}

We provided explicit expressions for the central charges and critical exponents in perturbation theory for a generic weakly coupled gauge-Yukawa theory. The conformal data are naturally divided into local quantities characterising a specific CFT and global data which are defined over the entire RG flow in between two CFTs belonging to the same underlying QFT. The local quantities are critical exponents, central charges and the ratio $a/c$ of two central charges. The variation of the central a-charge over the RG group defines the globally defined quantity. We characterised via their conformal data a wide class of fundamental (i.e. either free or safe) nonsupersymmetric quantum field theories dynamically developing CFTs at the end points of their perturbative RG flows. These theories are both vector and chiral like and constitute the backbone of phenomenologically interesting fundamental extensions of the Standard Model. Additionally our results can also be used as independent tests of the perturbative control over the possible existence of CFTs. 
We noted that the positivity of $a$ and the conformal collider bound \eqref{eq:ColBounds} are the most sensitive criteria. In contrast the positivity of the $c$ charge doesn't provide strong constraints (similar observation was made in the supersymmetric case  in \cite{Martin:2000cr}). Interestingly we show that at leading orders in perturbation theory the global variation of the $a$ charge is proportional to the critical exponent related to the relevant direction. This means that to this order one has one less independent conformal data.  Moreover, we extended this result beyond the cases in which the specific CFTs are achieved perturbatively by making use of conformal perturbation theory provided the two CFTs are nearby in coupling space.  Using  $1/N_f$ resummation techniques we also constructed an explicit example in which a strongly coupled safe CFT emerges that can be investigated using conformal pertrubation theory for which we can determine $\Delta a$. Interestingly this theory features an Higgs-like state and constitutes the template on which novel asymptotically safe Standard Models extensions have been constructed \cite{Mann:2017wzh,Abel:2017rwl,Pelaggi:2017abg,Antipin:2017ebo}.

\subsection*{Acknowledgements}
The CP$^3$-Origins centre is partially funded by the Danish National Research Foundation, grant number DNRF90. We thank Kasper Lang\ae ble, Esben M\o lgaard, Colin Poole and Zhi-Wei Wang for useful discussions. VP would like to thank CP$^3$-Origins for hospitality during initial and final stages of this work. VP was supported by the ERC STG grant 639220 (curvedsusy) during final stages of the project.

\appendix

\section{Perturbative a-theorem}
\label{3loops} 

The matrix element on the vacuum state of the trace of the energy--momentum tensor for the metric  $\gamma_{\mu \nu}$ for a general QFT in $d=4$ reads
\begin{equation}
	\left\langle T_\mu^\mu \right\rangle = c \, W^2(\gamma_{\mu\nu})
		- a \, E_4(\gamma_{\mu\nu}) + \ldots \ ,
\end{equation}
where $a$ and $c$ are real coefficients, $E_4(\gamma_{\mu\nu})$ the Euler density and $ W(\gamma_{\mu\nu})$ the Weyl tensor. The dots represent contributions coming from operators that can be constructed out of the fields defining the theory. Their contribution is proportional to the \betafunctions of their couplings. The coefficient $a$ is the one used in Cardy's conjecture, and for a free field theory it is~\cite{Duff:1977ay}
\begin{equation}
	\afree = \frac{1}{90 (8\pi)^2} \left( n_{\phi} + \frac{11}{2} n_{\psi} + 62 n_v \right) \ ,
	\label{eq:afree}
\end{equation}
where $n_{\psi}$, $n_{\psi}$ and $n_v$ are respectively the number of real scalars, Weyl fermions and gauge bosons. 

The change of $a$ along the RG flow is directly related to the underlying dynamics of the theory via the \betafunctions. This can be shown by exploiting the abelian nature of the trace anomaly which leads to the Weyl consistency conditions in much the same manner as the well known Wess-Zumino consistency conditions~\cite{Wess:1971yu}. Following the work of Jack and Osborn~\cite{Osborn:1989td, Jack:1990eb},  rather than using $a$ one uses the function $\tilde{a}$ related to it by
\begin{equation}
	\tilde{a} = a + w_i \beta_i \ ,
	\label{eq:atilde}
\end{equation}
where $w_i$ is a one--form which depends on the couplings of the theory. The Weyl consistency conditions imply for $\tilde{a}$ 
\begin{equation}
	\partial_i \tilde{a} = -\chi_{ij} \beta^j + (\partial_i w_j - \partial_j w_i) \beta^j \ ,
	\label{eq:consistencycondition}
\end{equation}
where $\chi_{ij}$ can be viewed as a metric in the space of couplings $g_i$. The positivity of the metric $\chi$ is established in perturbation theory, and therefore in this regime the function $\tilde{a}$ is monotonic along the RG flow
\begin{equation}
	\mu \frac{d\tilde{a}}{d\mu} = \beta^i \partial_i \tilde{a}
		= -\chi_{ij} \beta^i \beta^j \leq 0 \ .
\end{equation}
The irreversibility of the RG flow has been conjectured to be valid beyond perturbation theory at least in $d=4$.
%
%
  
For a generic gauge-Yukawa theory, the function $w^i$ turns out to be an exact one-form at the lowest orders in perturbation theory \cite{Jack:1990eb}, so that the terms involving derivatives of $w^i$ cancel out, and we will use in the following the simplified consistency condition
\begin{equation}
	\frac{\partial \tilde{a}}{\partial g^i} = -\beta_i \, ,
	\hspace{2cm}
	\beta_i \equiv \chi_{ij} \beta^j.
	\label{eq:truncatedconsistencycondition}
\end{equation}
$\chi_{ij}$ can be seen as a metric in the space of couplings, used in this case to raise and lower latin-indices. The fact that all \betafunctions can be derived from the same quantity $\tilde{a}$ has profound implications. The flow generated by the modified \betafunctions $\beta_i$ is a gradient flow, implying in particular
\begin{equation}
	\frac{\partial \beta_j}{\partial g^i} = \frac{\partial \beta_i}{\partial g^j} \, ,
	\label{eq:integrabilitycondition}
\end{equation}
which gives relations between the \betafunctions of different couplings. In the case of coupling-independent and diagonal metric, this relation reduces to
\begin{equation} \label{eq:integCondConst}
\chi_{ii}\partial_j \beta^i = \chi_{jj} \partial_i \beta^j \, ,
\end{equation}
where no summation on $i,j$ is present. \\ 
These consistency conditions, known as Weyl consistency conditions can be used as a check of a known computation. In principle they can also be used to determine some unknown coefficients at a higher loop order in perturbation theory.

\subsection{$\tilde{a}$ beyond two loops}
\label{sec:beyond2loops}

Is it possible to go beyond the two-loop calculation for the $a$ function previously introduced. For simplicity we consider a theory with one gauge group, one Yukawa interaction and two couplings in the scalar potential. In this case we can express our result in terms of $\alpha_i = g_i^2/(4\pi)^2$ to get a more compact notation. We will consider the scalar to be charged under the gauge group, generalizing the result in \cite{Antipin:2013pya}.\\
The theory is described by the general $\beta$-function system
\begin{eqnarray}
	\beta_{\alpha_g}/(-2 \alpha_g^2) &=& b_0 + b_1 \alpha_g + b_y \alpha_y 
		+b_3 \alpha_g^2 + b_4 \alpha_g \alpha_y + b_5 \alpha_y^2 + b_6 \alpha_{\lambda1} + b_7 \alpha_{\lambda1} \alpha_g + b_8 \alpha_{\lambda1}^2 \\
		&+& b_9 \alpha_{\lambda2} + b_{10} \alpha_{\lambda2} \alpha_g + b_{11} \alpha_{\lambda2}^2  + b_{12} \alpha_{\lambda1}\alpha_{\lambda2} \ ,
	\label{eq:betagauge} \\
	\beta_{\alpha_y}/(2\alpha_y)  &=&  c_1 \alpha_y + c_2 \alpha_g
		+ c_3 \alpha_g \alpha_y + c_4 \alpha_g^2 + c_5 \alpha_y^2
		+ c_6 \alpha_y \alpha_{\lambda1} + c_7 \alpha_{\lambda1}^2 + c_8 \alpha_{\lambda1} \alpha_g \\
		&+& c_{9} \alpha_y \alpha_{\lambda_2} + c_{10} \alpha_{\lambda_2}^2 + c_{11} \alpha_{\lambda_2} \alpha_g + c_{12} \alpha_{\lambda_1}\alpha_{\lambda_2} \ ,
	\label{eq:betaYukawa} \\ \notag
	\beta_{\alpha_{\lambda1}} & =&  d_1 \alpha_{\lambda1}^2+d_2 \alpha_{\lambda1} \alpha_y +d_3 \alpha_y^2+ d_4 \alpha_g^2 + d_5 \alpha_g \alpha_{\lambda1} + d_6  \alpha_{\lambda_2}^2 + d_7 \alpha_y \alpha_{\lambda_2}  \\
	& &+ d_8 \alpha_g \alpha_{\lambda_2} + d_9 \alpha_{\lambda_1} \alpha_{\lambda_2} \ .
	\label{eq:betaquartic1} \\ \notag
		\beta_{\alpha_{\lambda_2}} & = &  e_1 \alpha_{\lambda_2}^2+e_2 \alpha_{\lambda_2} \alpha_y +e_3 \alpha_y^2+ e_4 \alpha_g^2 + e_5 \alpha_g \alpha_{\lambda_2} + e_6  \alpha_{\lambda_1}^2 + e_7 \alpha_y \alpha_{\lambda_1} \\
		& & + e_8 \alpha_g \alpha_{\lambda_1} + e_9 \alpha_{\lambda_1} \alpha_{\lambda_2} \ .
	\label{eq:betaquartic2}
\end{eqnarray}
The Zamolodchidov metric is then generalized to
\begin{equation}
	\chi =
	\left(\begin{array}{cccc}
		\frac{\chi_{gg}}{\alpha_g^2} \left(1 + A \alpha_g + B_1 \alpha_g^2 + b_y \alpha_g \alpha_y + B_5 \alpha_g \alpha_{\lambda_1}+ B_6 \alpha_g \alpha_{\lambda_2} \right) & B_0 & E_0 & F_0 \\
		B_0 & \frac{\chi_{yy}}{\alpha_y} \left( 1 + B_3 \alpha_y + B_4 \alpha_g \right) & 0 & 0\\
		E_0 & 0 & \chi_{\lambda_1\lambda_1} & F_1 \\
		F_0 & 0 & F_1 & \chi_{\lambda_2\lambda_2} \\
	\end{array}\right) \ .
	\label{eq:metric}
\end{equation}
The coefficient $\chi_{gg}$ enters at the one-loop order, $A$ and $\chi_{yy}$ at two loops, while $\chi_{\lambda\lambda}$ and the $B_i$'s appear only at three loops. New non zero mixing terms are introduced in accordance with the definition of $\chi$ as a function of the 2-point function of stress-energy tensor. Similarly, the one-form $W$ reads
\begin{eqnarray}
	W_g & = & \frac{1}{\alpha_g} \left(D_0 + D_1 \alpha_g 
		+ C_1 \alpha_g^2 + C_2 \alpha_g \alpha_y  + C_5 \alpha_g \alpha_{\lambda_1} + C_6 \alpha_g \alpha_{\lambda_2} \right) \ ,\nonumber \\
	W_y & = & D_2 + C_3 \alpha_y + C_4 \alpha_g \ ,  \label{eq:oneform} \\
	W_{\lambda_1} & = & D_3 \alpha_{\lambda_1} + D_4 \alpha_g + D_5 \alpha_{\lambda_2} \nonumber \ . \\
	W_{\lambda_2} & = & E_3 \alpha_{\lambda_2} + E_4 \alpha_g + E_5 \alpha_{\lambda_1} \nonumber \ .
\end{eqnarray}
The general structure of $\chi$ confirms that it is sufficient for all our purposes to consider the Yukawa \betafunction (\ref{eq:betaYukawa}) to two-loop order and the quartic one (\ref{eq:betaquartic1}),(\ref{eq:betaquartic2})  to one-loop only. \\
The $\tilde{a}$ function is then derived up to three loop order
\begin{equation}\begin{split}
	\tilde{a} (\alpha_i ) = & - \frac{1}{3}\, \chi_{gg}\Big[ 4 b_0\, \alpha_g + \alpha_2^{2} (b_1 - 3\, A\,b_y)+ 2 A\, b_1 \alpha_g^3 + \left( b_4 +3 A b_y \right) \alpha_g^2 \alpha_y + 4 b_5\, \alpha_g \alpha_y^2 + \frac{c_1}{c_2} \left( 4 b_5-b_4  \frac{c_1}{c_2} \right) \alpha_y^3 - \frac{1}{\alpha_g}\beta_{\alpha_g} \Big] \nonumber \\
		&  +\frac{1}{3}\chi_{yy} \left[ \frac{1}{3} c_1 \alpha_y^2 + \frac{2}{3} c_2 \alpha_g \alpha_y -  \left( 2\, \left(\frac{c_1}{c_2}\right) c_3
			+\frac{1}{2} \left( \frac{c_1}{c_2} \right)^2 \left(A c_2 - 2 c_4\right) \right) \alpha_y^3
		- 2 c_3 \alpha_y^2 \alpha_{g}- 2 c_4 \alpha_y \alpha_{g}^2 +\beta_{\alpha_y} \right] \\
	&	+ \frac{1}{3}  \chi_{\lambda_1\lambda_1} \alpha_{\lambda_1} \beta_{\alpha_{\lambda_1}}+  \frac{1}{3}  \chi_{\lambda_2\lambda_2} \alpha_{\lambda_2} \beta_{\alpha_{\lambda_2}} + \frac{\beta_{\alpha_g}}{\alpha_g^2} f\left( \alpha_i^3 \right)- \frac{\beta_{\alpha_y}^2}{4\, \alpha_y} \frac{B_0 - C_2 + C_4}{c_2}\\
	& + \frac{1}{3}  (D_5 - E_5 + F_1) \alpha_{\lambda_2} \beta_{\alpha_{\lambda_1}}- \frac{1}{3} (D_5 - E_5 - F_1) \alpha_{\lambda_1} \beta_{\alpha_{\lambda_2}} \nonumber
\end{split}\end{equation}
where we defined
\begin{equation}\begin{split}
f\left( \alpha_i^3 \right) &=  \frac{2}{3}E_0 \alpha_g^2 \alpha_{\lambda1} +\frac{2}{3}F_0 \alpha_g^2 \alpha_{\lambda2} + \chi_{gg} \left( \frac{B_1}{3} \alpha_g^3 + \frac{b_y}{2} \alpha_g^2 \alpha_y - \frac{b_y}{6} \left( \frac{c_1}{c_2} \right)^2 \alpha_y^3 + \frac{B_5}{3}\alpha_g^2 \alpha_{\lambda_1} + \frac{B_6}{3}\alpha_g^2 \alpha_{\lambda_2}\right)\\
& + \frac{B_0 + C_2 - C_4}{3} \left( \frac{c_1}{c_2} \right)^2 \alpha_y^3 \\
\end{split}\end{equation}
the a-function coincides with this expression evaluated at a fixed point. As expected, all off diagonal terms in the metric and coefficients of the unphysical quantity $W$ drop at fixed point. Additionally all the dependence on scalar dynamics drops at a fixed point; every scalar beta function coefficient can be rearranged to form a beta function term using the Weyl consistency conditions. This is not accidental: it happens also at 2-1-0 level for the Yukawa coefficients, as shown in the previous section. We can then conjecture that the sector that is evaluated at the lowest order in perturbation theory does not explicitly contribute to the $a$-function. Nevertheless, it will still contribute to the fixed point value of the gauge and Yukawa couplings in this case.

\section{Conformal perturbation theory }

In this set-up we will consider a (strongly-)coupled CFT, which is described by the set of nearly marginal primary operators $O_i$ with small anomalous dimensions (critical exponents) $\Delta_i= 4+ \theta_{(i)}$ and the corresponding OPE (operator product expansion)
\begin{equation}
O_i (y) O_j(x) \stackrel{x \to y}{=} \frac{c_{ij}^k}{|x-y|^{\Delta_i+\Delta_j- \Delta_k}} O_k(x) + \dots \; ,
\end{equation} 
where dots correspond to operators of higher spin etc. We now proceed to deform the CFT by adding weakly relevant couplings $\{\lambda^i\}$
\begin{equation}
S_{\text{CFT}} \to S_{\text{CFT}} + \lambda^i \int d^4x O_i \; .
\end{equation}
Assuming small perturbations with $|\lambda^i| \ll 1$ we will have the corresponding beta functions
\begin{equation} \label{eq:CPTbetas}
\beta^i = \theta_{(i)} \lambda^i + \pi^2 \sum_{jk} c_{jk}^i \lambda^j \lambda^k + \mathcal{O}(\lambda^3) \; .
\end{equation}
 This reasoning can be reversed to obtain the conformal data from the knowledge of beta functions close to a fixed point as was done for example in \cite{Codello:2017hhh, Codello:2017epp}
\begin{equation}
\theta_{(i)}= \frac{\partial}{\partial \lambda^i} \beta^i |_{\lambda^{i*}} \; ; \quad  c_{jk}^i = \frac{1}{2 \pi^2 } \frac{\partial^2}{\partial \lambda^k \partial\lambda^j} \beta^i |_{\lambda^{i*}}    \; .
\end{equation} 
It should be noted that the above relation to compute OPE coefficients is strictly speaking only valid in the diagonal basis (for a more detailed discussion of this issue see section 2.3 in \cite{Codello:2017hhh}). 
For nearly marginal flows with small $\theta_{(i)}$, there is a possibility of Wilson-Fisher-like IR fixed point with small $\lambda^{i*}$.

\bibliographystyle{apsrev4-1}
\bibliography{ConformalDataFinal}

\begin{thebibliography}{53}%
\makeatletter
\providecommand \@ifxundefined [1]{%
 \@ifx{#1\undefined}
}%
\providecommand \@ifnum [1]{%
 \ifnum #1\expandafter \@firstoftwo
 \else \expandafter \@secondoftwo
 \fi
}%
\providecommand \@ifx [1]{%
 \ifx #1\expandafter \@firstoftwo
 \else \expandafter \@secondoftwo
 \fi
}%
\providecommand \natexlab [1]{#1}%
\providecommand \enquote  [1]{``#1''}%
\providecommand \bibnamefont  [1]{#1}%
\providecommand \bibfnamefont [1]{#1}%
\providecommand \citenamefont [1]{#1}%
\providecommand \href@noop [0]{\@secondoftwo}%
\providecommand \href [0]{\begingroup \@sanitize@url \@href}%
\providecommand \@href[1]{\@@startlink{#1}\@@href}%
\providecommand \@@href[1]{\endgroup#1\@@endlink}%
\providecommand \@sanitize@url [0]{\catcode `\\12\catcode `\$12\catcode
  `\&12\catcode `\#12\catcode `\^12\catcode `\_12\catcode `\%12\relax}%
\providecommand \@@startlink[1]{}%
\providecommand \@@endlink[0]{}%
\providecommand \url  [0]{\begingroup\@sanitize@url \@url }%
\providecommand \@url [1]{\endgroup\@href {#1}{\urlprefix }}%
\providecommand \urlprefix  [0]{URL }%
\providecommand \Eprint [0]{\href }%
\providecommand \doibase [0]{http://dx.doi.org/}%
\providecommand \selectlanguage [0]{\@gobble}%
\providecommand \bibinfo  [0]{\@secondoftwo}%
\providecommand \bibfield  [0]{\@secondoftwo}%
\providecommand \translation [1]{[#1]}%
\providecommand \BibitemOpen [0]{}%
\providecommand \bibitemStop [0]{}%
\providecommand \bibitemNoStop [0]{.\EOS\space}%
\providecommand \EOS [0]{\spacefactor3000\relax}%
\providecommand \BibitemShut  [1]{\csname bibitem#1\endcsname}%
\let\auto@bib@innerbib\@empty
\bibitem [{\citenamefont {Wilson}(1971{\natexlab{a}})}]{Wilson:1971bg}%
  \BibitemOpen
  \bibfield  {author} {\bibinfo {author} {\bibfnamefont {K.~G.}\ \bibnamefont
  {Wilson}},\ }\href {\doibase 10.1103/PhysRevB.4.3174} {\bibfield  {journal}
  {\bibinfo  {journal} {Phys. Rev.}\ }\textbf {\bibinfo {volume} {B4}},\
  \bibinfo {pages} {3174} (\bibinfo {year} {1971}{\natexlab{a}})}\BibitemShut
  {NoStop}%
\bibitem [{\citenamefont {Wilson}(1971{\natexlab{b}})}]{Wilson:1971dh}%
  \BibitemOpen
  \bibfield  {author} {\bibinfo {author} {\bibfnamefont {K.~G.}\ \bibnamefont
  {Wilson}},\ }\href {\doibase 10.1103/PhysRevB.4.3184} {\bibfield  {journal}
  {\bibinfo  {journal} {Phys. Rev.}\ }\textbf {\bibinfo {volume} {B4}},\
  \bibinfo {pages} {3184} (\bibinfo {year} {1971}{\natexlab{b}})}\BibitemShut
  {NoStop}%
\bibitem [{\citenamefont {Gross}\ and\ \citenamefont
  {Wilczek}(1973)}]{Gross:1973ju}%
  \BibitemOpen
  \bibfield  {author} {\bibinfo {author} {\bibfnamefont {D.~J.}\ \bibnamefont
  {Gross}}\ and\ \bibinfo {author} {\bibfnamefont {F.}~\bibnamefont
  {Wilczek}},\ }\href {\doibase 10.1103/PhysRevD.8.3633} {\bibfield  {journal}
  {\bibinfo  {journal} {Phys. Rev.}\ }\textbf {\bibinfo {volume} {D8}},\
  \bibinfo {pages} {3633} (\bibinfo {year} {1973})}\BibitemShut {NoStop}%
\bibitem [{\citenamefont {Politzer}(1973)}]{Politzer:1973fx}%
  \BibitemOpen
  \bibfield  {author} {\bibinfo {author} {\bibfnamefont {H.~D.}\ \bibnamefont
  {Politzer}},\ }\href {\doibase 10.1103/PhysRevLett.30.1346} {\bibfield
  {journal} {\bibinfo  {journal} {Phys. Rev. Lett.}\ }\textbf {\bibinfo
  {volume} {30}},\ \bibinfo {pages} {1346} (\bibinfo {year}
  {1973})}\BibitemShut {NoStop}%
\bibitem [{\citenamefont {Litim}\ and\ \citenamefont
  {Sannino}(2014)}]{Litim:2014uca}%
  \BibitemOpen
  \bibfield  {author} {\bibinfo {author} {\bibfnamefont {D.~F.}\ \bibnamefont
  {Litim}}\ and\ \bibinfo {author} {\bibfnamefont {F.}~\bibnamefont
  {Sannino}},\ }\href {\doibase 10.1007/JHEP12(2014)178} {\bibfield  {journal}
  {\bibinfo  {journal} {JHEP}\ }\textbf {\bibinfo {volume} {12}},\ \bibinfo
  {pages} {178} (\bibinfo {year} {2014})},\ \Eprint
  {http://arxiv.org/abs/1406.2337} {arXiv:1406.2337 [hep-th]} \BibitemShut
  {NoStop}%
\bibitem [{\citenamefont {Bond}\ and\ \citenamefont
  {Litim}(2018)}]{Bond:2017lnq}%
  \BibitemOpen
  \bibfield  {author} {\bibinfo {author} {\bibfnamefont {A.~D.}\ \bibnamefont
  {Bond}}\ and\ \bibinfo {author} {\bibfnamefont {D.~F.}\ \bibnamefont
  {Litim}},\ }\href {\doibase 10.1103/PhysRevD.97.085008} {\bibfield  {journal}
  {\bibinfo  {journal} {Phys. Rev.}\ }\textbf {\bibinfo {volume} {D97}},\
  \bibinfo {pages} {085008} (\bibinfo {year} {2018})},\ \Eprint
  {http://arxiv.org/abs/1707.04217} {arXiv:1707.04217 [hep-th]} \BibitemShut
  {NoStop}%
\bibitem [{\citenamefont {Bond}\ and\ \citenamefont
  {Litim}(2017)}]{Bond:2017suy}%
  \BibitemOpen
  \bibfield  {author} {\bibinfo {author} {\bibfnamefont {A.~D.}\ \bibnamefont
  {Bond}}\ and\ \bibinfo {author} {\bibfnamefont {D.~F.}\ \bibnamefont
  {Litim}},\ }\href {\doibase 10.1103/PhysRevLett.119.211601} {\bibfield
  {journal} {\bibinfo  {journal} {Phys. Rev. Lett.}\ }\textbf {\bibinfo
  {volume} {119}},\ \bibinfo {pages} {211601} (\bibinfo {year} {2017})},\
  \Eprint {http://arxiv.org/abs/1709.06953} {arXiv:1709.06953 [hep-th]}
  \BibitemShut {NoStop}%
\bibitem [{\citenamefont {Bajc}\ \emph {et~al.}(2018)\citenamefont {Bajc},
  \citenamefont {Dondi},\ and\ \citenamefont {Sannino}}]{Bajc:2017xwx}%
  \BibitemOpen
  \bibfield  {author} {\bibinfo {author} {\bibfnamefont {B.}~\bibnamefont
  {Bajc}}, \bibinfo {author} {\bibfnamefont {N.~A.}\ \bibnamefont {Dondi}}, \
  and\ \bibinfo {author} {\bibfnamefont {F.}~\bibnamefont {Sannino}},\ }\href
  {\doibase 10.1007/JHEP03(2018)005} {\bibfield  {journal} {\bibinfo  {journal}
  {JHEP}\ }\textbf {\bibinfo {volume} {03}},\ \bibinfo {pages} {005} (\bibinfo
  {year} {2018})},\ \Eprint {http://arxiv.org/abs/1709.07436} {arXiv:1709.07436
  [hep-th]} \BibitemShut {NoStop}%
\bibitem [{\citenamefont {Abel}\ and\ \citenamefont
  {Sannino}(2017{\natexlab{a}})}]{Abel:2017ujy}%
  \BibitemOpen
  \bibfield  {author} {\bibinfo {author} {\bibfnamefont {S.}~\bibnamefont
  {Abel}}\ and\ \bibinfo {author} {\bibfnamefont {F.}~\bibnamefont {Sannino}},\
  }\href {\doibase 10.1103/PhysRevD.96.056028} {\bibfield  {journal} {\bibinfo
  {journal} {Phys. Rev.}\ }\textbf {\bibinfo {volume} {D96}},\ \bibinfo {pages}
  {056028} (\bibinfo {year} {2017}{\natexlab{a}})},\ \Eprint
  {http://arxiv.org/abs/1704.00700} {arXiv:1704.00700 [hep-ph]} \BibitemShut
  {NoStop}%
\bibitem [{\citenamefont {Abel}\ and\ \citenamefont
  {Sannino}(2017{\natexlab{b}})}]{Abel:2017rwl}%
  \BibitemOpen
  \bibfield  {author} {\bibinfo {author} {\bibfnamefont {S.}~\bibnamefont
  {Abel}}\ and\ \bibinfo {author} {\bibfnamefont {F.}~\bibnamefont {Sannino}},\
  }\href {\doibase 10.1103/PhysRevD.96.055021} {\bibfield  {journal} {\bibinfo
  {journal} {Phys. Rev.}\ }\textbf {\bibinfo {volume} {D96}},\ \bibinfo {pages}
  {055021} (\bibinfo {year} {2017}{\natexlab{b}})},\ \Eprint
  {http://arxiv.org/abs/1707.06638} {arXiv:1707.06638 [hep-ph]} \BibitemShut
  {NoStop}%
\bibitem [{\citenamefont {Pelaggi}\ \emph
  {et~al.}(2017{\natexlab{a}})\citenamefont {Pelaggi}, \citenamefont {Sannino},
  \citenamefont {Strumia},\ and\ \citenamefont {Vigiani}}]{Pelaggi:2017wzr}%
  \BibitemOpen
  \bibfield  {author} {\bibinfo {author} {\bibfnamefont {G.~M.}\ \bibnamefont
  {Pelaggi}}, \bibinfo {author} {\bibfnamefont {F.}~\bibnamefont {Sannino}},
  \bibinfo {author} {\bibfnamefont {A.}~\bibnamefont {Strumia}}, \ and\
  \bibinfo {author} {\bibfnamefont {E.}~\bibnamefont {Vigiani}},\ }\href
  {\doibase 10.3389/fphy.2017.00049} {\bibfield  {journal} {\bibinfo  {journal}
  {Front.in Phys.}\ }\textbf {\bibinfo {volume} {5}},\ \bibinfo {pages} {49}
  (\bibinfo {year} {2017}{\natexlab{a}})},\ \Eprint
  {http://arxiv.org/abs/1701.01453} {arXiv:1701.01453 [hep-ph]} \BibitemShut
  {NoStop}%
\bibitem [{\citenamefont {Mann}\ \emph {et~al.}(2017)\citenamefont {Mann},
  \citenamefont {Meffe}, \citenamefont {Sannino}, \citenamefont {Steele},
  \citenamefont {Wang},\ and\ \citenamefont {Zhang}}]{Mann:2017wzh}%
  \BibitemOpen
  \bibfield  {author} {\bibinfo {author} {\bibfnamefont {R.}~\bibnamefont
  {Mann}}, \bibinfo {author} {\bibfnamefont {J.}~\bibnamefont {Meffe}},
  \bibinfo {author} {\bibfnamefont {F.}~\bibnamefont {Sannino}}, \bibinfo
  {author} {\bibfnamefont {T.}~\bibnamefont {Steele}}, \bibinfo {author}
  {\bibfnamefont {Z.-W.}\ \bibnamefont {Wang}}, \ and\ \bibinfo {author}
  {\bibfnamefont {C.}~\bibnamefont {Zhang}},\ }\href@noop {} {\  (\bibinfo
  {year} {2017})},\ \Eprint {http://arxiv.org/abs/1707.02942} {arXiv:1707.02942
  [hep-ph]} \BibitemShut {NoStop}%
\bibitem [{\citenamefont {Pelaggi}\ \emph
  {et~al.}(2017{\natexlab{b}})\citenamefont {Pelaggi}, \citenamefont
  {Plascencia}, \citenamefont {Salvio}, \citenamefont {Sannino}, \citenamefont
  {Smirnov},\ and\ \citenamefont {Strumia}}]{Pelaggi:2017abg}%
  \BibitemOpen
  \bibfield  {author} {\bibinfo {author} {\bibfnamefont {G.~M.}\ \bibnamefont
  {Pelaggi}}, \bibinfo {author} {\bibfnamefont {A.~D.}\ \bibnamefont
  {Plascencia}}, \bibinfo {author} {\bibfnamefont {A.}~\bibnamefont {Salvio}},
  \bibinfo {author} {\bibfnamefont {F.}~\bibnamefont {Sannino}}, \bibinfo
  {author} {\bibfnamefont {J.}~\bibnamefont {Smirnov}}, \ and\ \bibinfo
  {author} {\bibfnamefont {A.}~\bibnamefont {Strumia}},\ }\href@noop {} {\
  (\bibinfo {year} {2017}{\natexlab{b}})},\ \Eprint
  {http://arxiv.org/abs/1708.00437} {arXiv:1708.00437 [hep-ph]} \BibitemShut
  {NoStop}%
\bibitem [{\citenamefont {Bond}\ \emph {et~al.}(2017)\citenamefont {Bond},
  \citenamefont {Hiller}, \citenamefont {Kowalska},\ and\ \citenamefont
  {Litim}}]{Bond:2017wut}%
  \BibitemOpen
  \bibfield  {author} {\bibinfo {author} {\bibfnamefont {A.~D.}\ \bibnamefont
  {Bond}}, \bibinfo {author} {\bibfnamefont {G.}~\bibnamefont {Hiller}},
  \bibinfo {author} {\bibfnamefont {K.}~\bibnamefont {Kowalska}}, \ and\
  \bibinfo {author} {\bibfnamefont {D.~F.}\ \bibnamefont {Litim}},\ }\href
  {\doibase 10.1007/JHEP08(2017)004} {\bibfield  {journal} {\bibinfo  {journal}
  {JHEP}\ }\textbf {\bibinfo {volume} {08}},\ \bibinfo {pages} {004} (\bibinfo
  {year} {2017})},\ \Eprint {http://arxiv.org/abs/1702.01727} {arXiv:1702.01727
  [hep-ph]} \BibitemShut {NoStop}%
\bibitem [{\citenamefont {Antipin}\ and\ \citenamefont
  {Sannino}(2017)}]{Antipin:2017ebo}%
  \BibitemOpen
  \bibfield  {author} {\bibinfo {author} {\bibfnamefont {O.}~\bibnamefont
  {Antipin}}\ and\ \bibinfo {author} {\bibfnamefont {F.}~\bibnamefont
  {Sannino}},\ }\href@noop {} {\  (\bibinfo {year} {2017})},\ \Eprint
  {http://arxiv.org/abs/1709.02354} {arXiv:1709.02354 [hep-ph]} \BibitemShut
  {NoStop}%
\bibitem [{\citenamefont {Pica}\ and\ \citenamefont
  {Sannino}(2011)}]{Pica:2010xq}%
  \BibitemOpen
  \bibfield  {author} {\bibinfo {author} {\bibfnamefont {C.}~\bibnamefont
  {Pica}}\ and\ \bibinfo {author} {\bibfnamefont {F.}~\bibnamefont {Sannino}},\
  }\href {\doibase 10.1103/PhysRevD.83.035013} {\bibfield  {journal} {\bibinfo
  {journal} {Phys. Rev.}\ }\textbf {\bibinfo {volume} {D83}},\ \bibinfo {pages}
  {035013} (\bibinfo {year} {2011})},\ \Eprint {http://arxiv.org/abs/1011.5917}
  {arXiv:1011.5917 [hep-ph]} \BibitemShut {NoStop}%
\bibitem [{\citenamefont {Sannino}\ and\ \citenamefont
  {Tuominen}(2005)}]{Sannino:2004qp}%
  \BibitemOpen
  \bibfield  {author} {\bibinfo {author} {\bibfnamefont {F.}~\bibnamefont
  {Sannino}}\ and\ \bibinfo {author} {\bibfnamefont {K.}~\bibnamefont
  {Tuominen}},\ }\href {\doibase 10.1103/PhysRevD.71.051901} {\bibfield
  {journal} {\bibinfo  {journal} {Phys. Rev.}\ }\textbf {\bibinfo {volume}
  {D71}},\ \bibinfo {pages} {051901} (\bibinfo {year} {2005})},\ \Eprint
  {http://arxiv.org/abs/hep-ph/0405209} {arXiv:hep-ph/0405209 [hep-ph]}
  \BibitemShut {NoStop}%
\bibitem [{\citenamefont {Dietrich}\ and\ \citenamefont
  {Sannino}(2007)}]{Dietrich:2006cm}%
  \BibitemOpen
  \bibfield  {author} {\bibinfo {author} {\bibfnamefont {D.~D.}\ \bibnamefont
  {Dietrich}}\ and\ \bibinfo {author} {\bibfnamefont {F.}~\bibnamefont
  {Sannino}},\ }\href {\doibase 10.1103/PhysRevD.75.085018} {\bibfield
  {journal} {\bibinfo  {journal} {Phys. Rev.}\ }\textbf {\bibinfo {volume}
  {D75}},\ \bibinfo {pages} {085018} (\bibinfo {year} {2007})},\ \Eprint
  {http://arxiv.org/abs/hep-ph/0611341} {arXiv:hep-ph/0611341 [hep-ph]}
  \BibitemShut {NoStop}%
\bibitem [{\citenamefont {Jack}\ and\ \citenamefont
  {Poole}(2015)}]{Jack:2014pua}%
  \BibitemOpen
  \bibfield  {author} {\bibinfo {author} {\bibfnamefont {I.}~\bibnamefont
  {Jack}}\ and\ \bibinfo {author} {\bibfnamefont {C.}~\bibnamefont {Poole}},\
  }\href {\doibase 10.1007/JHEP01(2015)138} {\bibfield  {journal} {\bibinfo
  {journal} {JHEP}\ }\textbf {\bibinfo {volume} {01}},\ \bibinfo {pages} {138}
  (\bibinfo {year} {2015})},\ \Eprint {http://arxiv.org/abs/1411.1301}
  {arXiv:1411.1301 [hep-th]} \BibitemShut {NoStop}%
\bibitem [{\citenamefont {Antipin}\ \emph
  {et~al.}(2013{\natexlab{a}})\citenamefont {Antipin}, \citenamefont {Gillioz},
  \citenamefont {M{\o}lgaard},\ and\ \citenamefont
  {Sannino}}]{Antipin:2013pya}%
  \BibitemOpen
  \bibfield  {author} {\bibinfo {author} {\bibfnamefont {O.}~\bibnamefont
  {Antipin}}, \bibinfo {author} {\bibfnamefont {M.}~\bibnamefont {Gillioz}},
  \bibinfo {author} {\bibfnamefont {E.}~\bibnamefont {M{\o}lgaard}}, \ and\
  \bibinfo {author} {\bibfnamefont {F.}~\bibnamefont {Sannino}},\ }\href
  {\doibase 10.1103/PhysRevD.87.125017} {\bibfield  {journal} {\bibinfo
  {journal} {Phys. Rev.}\ }\textbf {\bibinfo {volume} {D87}},\ \bibinfo {pages}
  {125017} (\bibinfo {year} {2013}{\natexlab{a}})},\ \Eprint
  {http://arxiv.org/abs/1303.1525} {arXiv:1303.1525 [hep-th]} \BibitemShut
  {NoStop}%
\bibitem [{\citenamefont {Antipin}\ \emph
  {et~al.}(2013{\natexlab{b}})\citenamefont {Antipin}, \citenamefont {Gillioz},
  \citenamefont {Krog}, \citenamefont {M{\o}lgaard},\ and\ \citenamefont
  {Sannino}}]{Antipin:2013sga}%
  \BibitemOpen
  \bibfield  {author} {\bibinfo {author} {\bibfnamefont {O.}~\bibnamefont
  {Antipin}}, \bibinfo {author} {\bibfnamefont {M.}~\bibnamefont {Gillioz}},
  \bibinfo {author} {\bibfnamefont {J.}~\bibnamefont {Krog}}, \bibinfo {author}
  {\bibfnamefont {E.}~\bibnamefont {M{\o}lgaard}}, \ and\ \bibinfo {author}
  {\bibfnamefont {F.}~\bibnamefont {Sannino}},\ }\href {\doibase
  10.1007/JHEP08(2013)034} {\bibfield  {journal} {\bibinfo  {journal} {JHEP}\
  }\textbf {\bibinfo {volume} {08}},\ \bibinfo {pages} {034} (\bibinfo {year}
  {2013}{\natexlab{b}})},\ \Eprint {http://arxiv.org/abs/1306.3234}
  {arXiv:1306.3234 [hep-ph]} \BibitemShut {NoStop}%
\bibitem [{\citenamefont {Osborn}(1989)}]{Osborn:1989td}%
  \BibitemOpen
  \bibfield  {author} {\bibinfo {author} {\bibfnamefont {H.}~\bibnamefont
  {Osborn}},\ }\href {\doibase 10.1016/0370-2693(89)90729-6} {\bibfield
  {journal} {\bibinfo  {journal} {Phys. Lett.}\ }\textbf {\bibinfo {volume}
  {B222}},\ \bibinfo {pages} {97} (\bibinfo {year} {1989})}\BibitemShut
  {NoStop}%
\bibitem [{\citenamefont {Bond}\ \emph {et~al.}(2018)\citenamefont {Bond},
  \citenamefont {Litim}, \citenamefont {Medina~Vazquez},\ and\ \citenamefont
  {Steudtner}}]{Bond:2017tbw}%
  \BibitemOpen
  \bibfield  {author} {\bibinfo {author} {\bibfnamefont {A.~D.}\ \bibnamefont
  {Bond}}, \bibinfo {author} {\bibfnamefont {D.~F.}\ \bibnamefont {Litim}},
  \bibinfo {author} {\bibfnamefont {G.}~\bibnamefont {Medina~Vazquez}}, \ and\
  \bibinfo {author} {\bibfnamefont {T.}~\bibnamefont {Steudtner}},\ }\href
  {\doibase 10.1103/PhysRevD.97.036019} {\bibfield  {journal} {\bibinfo
  {journal} {Phys. Rev.}\ }\textbf {\bibinfo {volume} {D97}},\ \bibinfo {pages}
  {036019} (\bibinfo {year} {2018})},\ \Eprint
  {http://arxiv.org/abs/1710.07615} {arXiv:1710.07615 [hep-th]} \BibitemShut
  {NoStop}%
\bibitem [{\citenamefont {Jack}\ and\ \citenamefont
  {Poole}(2017)}]{Jack:2016utw}%
  \BibitemOpen
  \bibfield  {author} {\bibinfo {author} {\bibfnamefont {I.}~\bibnamefont
  {Jack}}\ and\ \bibinfo {author} {\bibfnamefont {C.}~\bibnamefont {Poole}},\
  }\href {\doibase 10.1103/PhysRevD.95.025010} {\bibfield  {journal} {\bibinfo
  {journal} {Phys. Rev.}\ }\textbf {\bibinfo {volume} {D95}},\ \bibinfo {pages}
  {025010} (\bibinfo {year} {2017})},\ \Eprint
  {http://arxiv.org/abs/1607.00236} {arXiv:1607.00236 [hep-th]} \BibitemShut
  {NoStop}%
\bibitem [{\citenamefont {Bednyakov}\ and\ \citenamefont
  {Pikelner}(2016)}]{Bednyakov:2015ooa}%
  \BibitemOpen
  \bibfield  {author} {\bibinfo {author} {\bibfnamefont {A.~V.}\ \bibnamefont
  {Bednyakov}}\ and\ \bibinfo {author} {\bibfnamefont {A.~F.}\ \bibnamefont
  {Pikelner}},\ }\href {\doibase 10.1016/j.physletb.2016.09.007} {\bibfield
  {journal} {\bibinfo  {journal} {Phys. Lett.}\ }\textbf {\bibinfo {volume}
  {B762}},\ \bibinfo {pages} {151} (\bibinfo {year} {2016})},\ \Eprint
  {http://arxiv.org/abs/1508.02680} {arXiv:1508.02680 [hep-ph]} \BibitemShut
  {NoStop}%
\bibitem [{\citenamefont {Jack}\ \emph {et~al.}(2015)\citenamefont {Jack},
  \citenamefont {Jones},\ and\ \citenamefont {Poole}}]{Jack:2015tka}%
  \BibitemOpen
  \bibfield  {author} {\bibinfo {author} {\bibfnamefont {I.}~\bibnamefont
  {Jack}}, \bibinfo {author} {\bibfnamefont {D.~R.~T.}\ \bibnamefont {Jones}},
  \ and\ \bibinfo {author} {\bibfnamefont {C.}~\bibnamefont {Poole}},\ }\href
  {\doibase 10.1007/JHEP09(2015)061} {\bibfield  {journal} {\bibinfo  {journal}
  {JHEP}\ }\textbf {\bibinfo {volume} {09}},\ \bibinfo {pages} {061} (\bibinfo
  {year} {2015})},\ \Eprint {http://arxiv.org/abs/1505.05400} {arXiv:1505.05400
  [hep-th]} \BibitemShut {NoStop}%
\bibitem [{\citenamefont {M\o~lgaard}(2014)}]{Molgaard:2014hpa}%
  \BibitemOpen
  \bibfield  {author} {\bibinfo {author} {\bibfnamefont {E.}~\bibnamefont
  {M\o~lgaard}},\ }\href {\doibase 10.1140/epjp/i2014-14159-2} {\bibfield
  {journal} {\bibinfo  {journal} {Eur. Phys. J. Plus}\ }\textbf {\bibinfo
  {volume} {129}},\ \bibinfo {pages} {159} (\bibinfo {year} {2014})},\ \Eprint
  {http://arxiv.org/abs/1404.5550} {arXiv:1404.5550 [hep-th]} \BibitemShut
  {NoStop}%
\bibitem [{\citenamefont {Jack}(1983)}]{Jack:1982sn}%
  \BibitemOpen
  \bibfield  {author} {\bibinfo {author} {\bibfnamefont {I.}~\bibnamefont
  {Jack}},\ }\href {\doibase 10.1088/0305-4470/16/5/025} {\bibfield  {journal}
  {\bibinfo  {journal} {J. Phys.}\ }\textbf {\bibinfo {volume} {A16}},\
  \bibinfo {pages} {1083} (\bibinfo {year} {1983})}\BibitemShut {NoStop}%
\bibitem [{\citenamefont {Osborn}\ and\ \citenamefont
  {Stergiou}(2016)}]{Osborn:2016bev}%
  \BibitemOpen
  \bibfield  {author} {\bibinfo {author} {\bibfnamefont {H.}~\bibnamefont
  {Osborn}}\ and\ \bibinfo {author} {\bibfnamefont {A.}~\bibnamefont
  {Stergiou}},\ }\href {\doibase 10.1007/JHEP06(2016)079} {\bibfield  {journal}
  {\bibinfo  {journal} {JHEP}\ }\textbf {\bibinfo {volume} {06}},\ \bibinfo
  {pages} {079} (\bibinfo {year} {2016})},\ \Eprint
  {http://arxiv.org/abs/1603.07307} {arXiv:1603.07307 [hep-th]} \BibitemShut
  {NoStop}%
\bibitem [{\citenamefont {Hofman}\ and\ \citenamefont
  {Maldacena}(2008)}]{Hofman:2008ar}%
  \BibitemOpen
  \bibfield  {author} {\bibinfo {author} {\bibfnamefont {D.~M.}\ \bibnamefont
  {Hofman}}\ and\ \bibinfo {author} {\bibfnamefont {J.}~\bibnamefont
  {Maldacena}},\ }\href {\doibase 10.1088/1126-6708/2008/05/012} {\bibfield
  {journal} {\bibinfo  {journal} {JHEP}\ }\textbf {\bibinfo {volume} {05}},\
  \bibinfo {pages} {012} (\bibinfo {year} {2008})},\ \Eprint
  {http://arxiv.org/abs/0803.1467} {arXiv:0803.1467 [hep-th]} \BibitemShut
  {NoStop}%
\bibitem [{\citenamefont {Parnachev}\ and\ \citenamefont
  {Razamat}(2009)}]{Parnachev:2008yt}%
  \BibitemOpen
  \bibfield  {author} {\bibinfo {author} {\bibfnamefont {A.}~\bibnamefont
  {Parnachev}}\ and\ \bibinfo {author} {\bibfnamefont {S.~S.}\ \bibnamefont
  {Razamat}},\ }\href {\doibase 10.1088/1126-6708/2009/07/010} {\bibfield
  {journal} {\bibinfo  {journal} {JHEP}\ }\textbf {\bibinfo {volume} {07}},\
  \bibinfo {pages} {010} (\bibinfo {year} {2009})},\ \Eprint
  {http://arxiv.org/abs/0812.0781} {arXiv:0812.0781 [hep-th]} \BibitemShut
  {NoStop}%
\bibitem [{\citenamefont {Komargodski}\ \emph {et~al.}(2017)\citenamefont
  {Komargodski}, \citenamefont {Kulaxizi}, \citenamefont {Parnachev},\ and\
  \citenamefont {Zhiboedov}}]{Komargodski:2016gci}%
  \BibitemOpen
  \bibfield  {author} {\bibinfo {author} {\bibfnamefont {Z.}~\bibnamefont
  {Komargodski}}, \bibinfo {author} {\bibfnamefont {M.}~\bibnamefont
  {Kulaxizi}}, \bibinfo {author} {\bibfnamefont {A.}~\bibnamefont {Parnachev}},
  \ and\ \bibinfo {author} {\bibfnamefont {A.}~\bibnamefont {Zhiboedov}},\
  }\href {\doibase 10.1103/PhysRevD.95.065011} {\bibfield  {journal} {\bibinfo
  {journal} {Phys. Rev.}\ }\textbf {\bibinfo {volume} {D95}},\ \bibinfo {pages}
  {065011} (\bibinfo {year} {2017})},\ \Eprint
  {http://arxiv.org/abs/1601.05453} {arXiv:1601.05453 [hep-th]} \BibitemShut
  {NoStop}%
\bibitem [{\citenamefont {Hofman}\ \emph {et~al.}(2016)\citenamefont {Hofman},
  \citenamefont {Li}, \citenamefont {Meltzer}, \citenamefont {Poland},\ and\
  \citenamefont {Rejon-Barrera}}]{Hofman:2016awc}%
  \BibitemOpen
  \bibfield  {author} {\bibinfo {author} {\bibfnamefont {D.~M.}\ \bibnamefont
  {Hofman}}, \bibinfo {author} {\bibfnamefont {D.}~\bibnamefont {Li}}, \bibinfo
  {author} {\bibfnamefont {D.}~\bibnamefont {Meltzer}}, \bibinfo {author}
  {\bibfnamefont {D.}~\bibnamefont {Poland}}, \ and\ \bibinfo {author}
  {\bibfnamefont {F.}~\bibnamefont {Rejon-Barrera}},\ }\href {\doibase
  10.1007/JHEP06(2016)111} {\bibfield  {journal} {\bibinfo  {journal} {JHEP}\
  }\textbf {\bibinfo {volume} {06}},\ \bibinfo {pages} {111} (\bibinfo {year}
  {2016})},\ \Eprint {http://arxiv.org/abs/1603.03771} {arXiv:1603.03771
  [hep-th]} \BibitemShut {NoStop}%
\bibitem [{\citenamefont {Komargodski}\ and\ \citenamefont
  {Schwimmer}(2011)}]{Komargodski:2011vj}%
  \BibitemOpen
  \bibfield  {author} {\bibinfo {author} {\bibfnamefont {Z.}~\bibnamefont
  {Komargodski}}\ and\ \bibinfo {author} {\bibfnamefont {A.}~\bibnamefont
  {Schwimmer}},\ }\href {\doibase 10.1007/JHEP12(2011)099} {\bibfield
  {journal} {\bibinfo  {journal} {JHEP}\ }\textbf {\bibinfo {volume} {12}},\
  \bibinfo {pages} {099} (\bibinfo {year} {2011})},\ \Eprint
  {http://arxiv.org/abs/1107.3987} {arXiv:1107.3987 [hep-th]} \BibitemShut
  {NoStop}%
\bibitem [{\citenamefont {Cappelli}\ \emph {et~al.}(1991)\citenamefont
  {Cappelli}, \citenamefont {Friedan},\ and\ \citenamefont
  {Latorre}}]{Cappelli:1990yc}%
  \BibitemOpen
  \bibfield  {author} {\bibinfo {author} {\bibfnamefont {A.}~\bibnamefont
  {Cappelli}}, \bibinfo {author} {\bibfnamefont {D.}~\bibnamefont {Friedan}}, \
  and\ \bibinfo {author} {\bibfnamefont {J.~I.}\ \bibnamefont {Latorre}},\
  }\href {\doibase 10.1016/0550-3213(91)90102-4} {\bibfield  {journal}
  {\bibinfo  {journal} {Nucl. Phys.}\ }\textbf {\bibinfo {volume} {B352}},\
  \bibinfo {pages} {616} (\bibinfo {year} {1991})}\BibitemShut {NoStop}%
\bibitem [{\citenamefont {Baume}\ \emph {et~al.}(2014)\citenamefont {Baume},
  \citenamefont {Keren-Zur}, \citenamefont {Rattazzi},\ and\ \citenamefont
  {Vitale}}]{Baume:2014rla}%
  \BibitemOpen
  \bibfield  {author} {\bibinfo {author} {\bibfnamefont {F.}~\bibnamefont
  {Baume}}, \bibinfo {author} {\bibfnamefont {B.}~\bibnamefont {Keren-Zur}},
  \bibinfo {author} {\bibfnamefont {R.}~\bibnamefont {Rattazzi}}, \ and\
  \bibinfo {author} {\bibfnamefont {L.}~\bibnamefont {Vitale}},\ }\href
  {\doibase 10.1007/JHEP08(2014)152} {\bibfield  {journal} {\bibinfo  {journal}
  {JHEP}\ }\textbf {\bibinfo {volume} {08}},\ \bibinfo {pages} {152} (\bibinfo
  {year} {2014})},\ \Eprint {http://arxiv.org/abs/1401.5983} {arXiv:1401.5983
  [hep-th]} \BibitemShut {NoStop}%
\bibitem [{\citenamefont {Jack}\ and\ \citenamefont
  {Osborn}(1990)}]{Jack:1990eb}%
  \BibitemOpen
  \bibfield  {author} {\bibinfo {author} {\bibfnamefont {I.}~\bibnamefont
  {Jack}}\ and\ \bibinfo {author} {\bibfnamefont {H.}~\bibnamefont {Osborn}},\
  }\href {\doibase 10.1016/0550-3213(90)90584-Z} {\bibfield  {journal}
  {\bibinfo  {journal} {Nucl. Phys.}\ }\textbf {\bibinfo {volume} {B343}},\
  \bibinfo {pages} {647} (\bibinfo {year} {1990})}\BibitemShut {NoStop}%
\bibitem [{\citenamefont {Auzzi}\ and\ \citenamefont
  {Keren-Zur}(2015)}]{Auzzi:2015yia}%
  \BibitemOpen
  \bibfield  {author} {\bibinfo {author} {\bibfnamefont {R.}~\bibnamefont
  {Auzzi}}\ and\ \bibinfo {author} {\bibfnamefont {B.}~\bibnamefont
  {Keren-Zur}},\ }\href {\doibase 10.1007/JHEP05(2015)150} {\bibfield
  {journal} {\bibinfo  {journal} {JHEP}\ }\textbf {\bibinfo {volume} {05}},\
  \bibinfo {pages} {150} (\bibinfo {year} {2015})},\ \Eprint
  {http://arxiv.org/abs/1502.05962} {arXiv:1502.05962 [hep-th]} \BibitemShut
  {NoStop}%
\bibitem [{\citenamefont {Gomis}\ \emph {et~al.}(2016)\citenamefont {Gomis},
  \citenamefont {Hsin}, \citenamefont {Komargodski}, \citenamefont {Schwimmer},
  \citenamefont {Seiberg},\ and\ \citenamefont {Theisen}}]{Gomis:2015yaa}%
  \BibitemOpen
  \bibfield  {author} {\bibinfo {author} {\bibfnamefont {J.}~\bibnamefont
  {Gomis}}, \bibinfo {author} {\bibfnamefont {P.-S.}\ \bibnamefont {Hsin}},
  \bibinfo {author} {\bibfnamefont {Z.}~\bibnamefont {Komargodski}}, \bibinfo
  {author} {\bibfnamefont {A.}~\bibnamefont {Schwimmer}}, \bibinfo {author}
  {\bibfnamefont {N.}~\bibnamefont {Seiberg}}, \ and\ \bibinfo {author}
  {\bibfnamefont {S.}~\bibnamefont {Theisen}},\ }\href {\doibase
  10.1007/JHEP03(2016)022} {\bibfield  {journal} {\bibinfo  {journal} {JHEP}\
  }\textbf {\bibinfo {volume} {03}},\ \bibinfo {pages} {022} (\bibinfo {year}
  {2016})},\ \Eprint {http://arxiv.org/abs/1509.08511} {arXiv:1509.08511
  [hep-th]} \BibitemShut {NoStop}%
\bibitem [{\citenamefont {Pica}\ \emph {et~al.}(2017)\citenamefont {Pica},
  \citenamefont {Ryttov},\ and\ \citenamefont {Sannino}}]{Pica:2016krb}%
  \BibitemOpen
  \bibfield  {author} {\bibinfo {author} {\bibfnamefont {C.}~\bibnamefont
  {Pica}}, \bibinfo {author} {\bibfnamefont {T.~A.}\ \bibnamefont {Ryttov}}, \
  and\ \bibinfo {author} {\bibfnamefont {F.}~\bibnamefont {Sannino}},\ }\href
  {\doibase 10.1103/PhysRevD.96.074015} {\bibfield  {journal} {\bibinfo
  {journal} {Phys. Rev.}\ }\textbf {\bibinfo {volume} {D96}},\ \bibinfo {pages}
  {074015} (\bibinfo {year} {2017})},\ \Eprint
  {http://arxiv.org/abs/1605.04712} {arXiv:1605.04712 [hep-th]} \BibitemShut
  {NoStop}%
\bibitem [{\citenamefont {Anselmi}\ \emph {et~al.}(1998)\citenamefont
  {Anselmi}, \citenamefont {Erlich}, \citenamefont {Freedman},\ and\
  \citenamefont {Johansen}}]{Anselmi:1997ys}%
  \BibitemOpen
  \bibfield  {author} {\bibinfo {author} {\bibfnamefont {D.}~\bibnamefont
  {Anselmi}}, \bibinfo {author} {\bibfnamefont {J.}~\bibnamefont {Erlich}},
  \bibinfo {author} {\bibfnamefont {D.~Z.}\ \bibnamefont {Freedman}}, \ and\
  \bibinfo {author} {\bibfnamefont {A.~A.}\ \bibnamefont {Johansen}},\ }\href
  {\doibase 10.1103/PhysRevD.57.7570} {\bibfield  {journal} {\bibinfo
  {journal} {Phys. Rev.}\ }\textbf {\bibinfo {volume} {D57}},\ \bibinfo {pages}
  {7570} (\bibinfo {year} {1998})},\ \Eprint
  {http://arxiv.org/abs/hep-th/9711035} {arXiv:hep-th/9711035 [hep-th]}
  \BibitemShut {NoStop}%
\bibitem [{\citenamefont {Prochazka}\ and\ \citenamefont
  {Zwicky}(2017)}]{Prochazka:2017pfa}%
  \BibitemOpen
  \bibfield  {author} {\bibinfo {author} {\bibfnamefont {V.}~\bibnamefont
  {Prochazka}}\ and\ \bibinfo {author} {\bibfnamefont {R.}~\bibnamefont
  {Zwicky}},\ }\href {\doibase 10.1103/PhysRevD.96.045011} {\bibfield
  {journal} {\bibinfo  {journal} {Phys. Rev.}\ }\textbf {\bibinfo {volume}
  {D96}},\ \bibinfo {pages} {045011} (\bibinfo {year} {2017})},\ \Eprint
  {http://arxiv.org/abs/1703.01239} {arXiv:1703.01239 [hep-th]} \BibitemShut
  {NoStop}%
\bibitem [{\citenamefont {Hansen}\ \emph {et~al.}(2017)\citenamefont {Hansen},
  \citenamefont {Janowski}, \citenamefont {Langaeble}, \citenamefont {Mann},
  \citenamefont {Sannino}, \citenamefont {Steele},\ and\ \citenamefont
  {Wang}}]{Hansen:2017pwe}%
  \BibitemOpen
  \bibfield  {author} {\bibinfo {author} {\bibfnamefont {F.~F.}\ \bibnamefont
  {Hansen}}, \bibinfo {author} {\bibfnamefont {T.}~\bibnamefont {Janowski}},
  \bibinfo {author} {\bibfnamefont {K.}~\bibnamefont {Langaeble}}, \bibinfo
  {author} {\bibfnamefont {R.~B.}\ \bibnamefont {Mann}}, \bibinfo {author}
  {\bibfnamefont {F.}~\bibnamefont {Sannino}}, \bibinfo {author} {\bibfnamefont
  {T.~G.}\ \bibnamefont {Steele}}, \ and\ \bibinfo {author} {\bibfnamefont
  {Z.-W.}\ \bibnamefont {Wang}},\ }\href@noop {} {\  (\bibinfo {year}
  {2017})},\ \Eprint {http://arxiv.org/abs/1706.06402} {arXiv:1706.06402
  [hep-ph]} \BibitemShut {NoStop}%
\bibitem [{\citenamefont {Appelquist}\ \emph {et~al.}(2000)\citenamefont
  {Appelquist}, \citenamefont {Duan},\ and\ \citenamefont
  {Sannino}}]{Appelquist:2000qg}%
  \BibitemOpen
  \bibfield  {author} {\bibinfo {author} {\bibfnamefont {T.}~\bibnamefont
  {Appelquist}}, \bibinfo {author} {\bibfnamefont {Z.-y.}\ \bibnamefont
  {Duan}}, \ and\ \bibinfo {author} {\bibfnamefont {F.}~\bibnamefont
  {Sannino}},\ }\href {\doibase 10.1103/PhysRevD.61.125009} {\bibfield
  {journal} {\bibinfo  {journal} {Phys. Rev.}\ }\textbf {\bibinfo {volume}
  {D61}},\ \bibinfo {pages} {125009} (\bibinfo {year} {2000})},\ \Eprint
  {http://arxiv.org/abs/hep-ph/0001043} {arXiv:hep-ph/0001043 [hep-ph]}
  \BibitemShut {NoStop}%
\bibitem [{\citenamefont {M{\o}lgaard}\ and\ \citenamefont
  {Sannino}(2017)}]{Molgaard:2016bqf}%
  \BibitemOpen
  \bibfield  {author} {\bibinfo {author} {\bibfnamefont {E.}~\bibnamefont
  {M{\o}lgaard}}\ and\ \bibinfo {author} {\bibfnamefont {F.}~\bibnamefont
  {Sannino}},\ }\href {\doibase 10.1103/PhysRevD.96.056004} {\bibfield
  {journal} {\bibinfo  {journal} {Phys. Rev.}\ }\textbf {\bibinfo {volume}
  {D96}},\ \bibinfo {pages} {056004} (\bibinfo {year} {2017})},\ \Eprint
  {http://arxiv.org/abs/1610.03130} {arXiv:1610.03130 [hep-ph]} \BibitemShut
  {NoStop}%
\bibitem [{\citenamefont {Georgi}\ \emph {et~al.}(1974)\citenamefont {Georgi},
  \citenamefont {Quinn},\ and\ \citenamefont {Weinberg}}]{PhysRevLett.33.451}%
  \BibitemOpen
  \bibfield  {author} {\bibinfo {author} {\bibfnamefont {H.}~\bibnamefont
  {Georgi}}, \bibinfo {author} {\bibfnamefont {H.~R.}\ \bibnamefont {Quinn}}, \
  and\ \bibinfo {author} {\bibfnamefont {S.}~\bibnamefont {Weinberg}},\ }\href
  {\doibase 10.1103/PhysRevLett.33.451} {\bibfield  {journal} {\bibinfo
  {journal} {Phys. Rev. Lett.}\ }\textbf {\bibinfo {volume} {33}},\ \bibinfo
  {pages} {451} (\bibinfo {year} {1974})}\BibitemShut {NoStop}%
\bibitem [{\citenamefont {Bars}\ and\ \citenamefont
  {Yankielowicz}(1981)}]{BARS1981159}%
  \BibitemOpen
  \bibfield  {author} {\bibinfo {author} {\bibfnamefont {I.}~\bibnamefont
  {Bars}}\ and\ \bibinfo {author} {\bibfnamefont {S.}~\bibnamefont
  {Yankielowicz}},\ }\href {\doibase
  https://doi.org/10.1016/0370-2693(81)90664-X} {\bibfield  {journal} {\bibinfo
   {journal} {Physics Letters B}\ }\textbf {\bibinfo {volume} {101}},\ \bibinfo
  {pages} {159 } (\bibinfo {year} {1981})}\BibitemShut {NoStop}%
\bibitem [{\citenamefont {Holdom}(2011)}]{Holdom:2010qs}%
  \BibitemOpen
  \bibfield  {author} {\bibinfo {author} {\bibfnamefont {B.}~\bibnamefont
  {Holdom}},\ }\href {\doibase 10.1016/j.physletb.2010.09.037} {\bibfield
  {journal} {\bibinfo  {journal} {Phys. Lett.}\ }\textbf {\bibinfo {volume}
  {B694}},\ \bibinfo {pages} {74} (\bibinfo {year} {2011})},\ \Eprint
  {http://arxiv.org/abs/1006.2119} {arXiv:1006.2119 [hep-ph]} \BibitemShut
  {NoStop}%
\bibitem [{\citenamefont {Martin}\ and\ \citenamefont
  {Wells}(2001)}]{Martin:2000cr}%
  \BibitemOpen
  \bibfield  {author} {\bibinfo {author} {\bibfnamefont {S.~P.}\ \bibnamefont
  {Martin}}\ and\ \bibinfo {author} {\bibfnamefont {J.~D.}\ \bibnamefont
  {Wells}},\ }\href {\doibase 10.1103/PhysRevD.64.036010} {\bibfield  {journal}
  {\bibinfo  {journal} {Phys. Rev.}\ }\textbf {\bibinfo {volume} {D64}},\
  \bibinfo {pages} {036010} (\bibinfo {year} {2001})},\ \Eprint
  {http://arxiv.org/abs/hep-ph/0011382} {arXiv:hep-ph/0011382 [hep-ph]}
  \BibitemShut {NoStop}%
\bibitem [{\citenamefont {Duff}(1977)}]{Duff:1977ay}%
  \BibitemOpen
  \bibfield  {author} {\bibinfo {author} {\bibfnamefont {M.~J.}\ \bibnamefont
  {Duff}},\ }\href {\doibase 10.1016/0550-3213(77)90410-2} {\bibfield
  {journal} {\bibinfo  {journal} {Nucl. Phys.}\ }\textbf {\bibinfo {volume}
  {B125}},\ \bibinfo {pages} {334} (\bibinfo {year} {1977})}\BibitemShut
  {NoStop}%
\bibitem [{\citenamefont {Wess}\ and\ \citenamefont
  {Zumino}(1971)}]{Wess:1971yu}%
  \BibitemOpen
  \bibfield  {author} {\bibinfo {author} {\bibfnamefont {J.}~\bibnamefont
  {Wess}}\ and\ \bibinfo {author} {\bibfnamefont {B.}~\bibnamefont {Zumino}},\
  }\href {\doibase 10.1016/0370-2693(71)90582-X} {\bibfield  {journal}
  {\bibinfo  {journal} {Phys. Lett.}\ }\textbf {\bibinfo {volume} {37B}},\
  \bibinfo {pages} {95} (\bibinfo {year} {1971})}\BibitemShut {NoStop}%
\bibitem [{\citenamefont {Codello}\ \emph {et~al.}(2018)\citenamefont
  {Codello}, \citenamefont {Safari}, \citenamefont {Vacca},\ and\ \citenamefont
  {Zanusso}}]{Codello:2017hhh}%
  \BibitemOpen
  \bibfield  {author} {\bibinfo {author} {\bibfnamefont {A.}~\bibnamefont
  {Codello}}, \bibinfo {author} {\bibfnamefont {M.}~\bibnamefont {Safari}},
  \bibinfo {author} {\bibfnamefont {G.~P.}\ \bibnamefont {Vacca}}, \ and\
  \bibinfo {author} {\bibfnamefont {O.}~\bibnamefont {Zanusso}},\ }\href
  {\doibase 10.1140/epjc/s10052-017-5505-2} {\bibfield  {journal} {\bibinfo
  {journal} {Eur. Phys. J.}\ }\textbf {\bibinfo {volume} {C78}},\ \bibinfo
  {pages} {30} (\bibinfo {year} {2018})},\ \Eprint
  {http://arxiv.org/abs/1705.05558} {arXiv:1705.05558 [hep-th]} \BibitemShut
  {NoStop}%
\bibitem [{\citenamefont {Codello}\ \emph {et~al.}(2017)\citenamefont
  {Codello}, \citenamefont {Safari}, \citenamefont {Vacca},\ and\ \citenamefont
  {Zanusso}}]{Codello:2017epp}%
  \BibitemOpen
  \bibfield  {author} {\bibinfo {author} {\bibfnamefont {A.}~\bibnamefont
  {Codello}}, \bibinfo {author} {\bibfnamefont {M.}~\bibnamefont {Safari}},
  \bibinfo {author} {\bibfnamefont {G.~P.}\ \bibnamefont {Vacca}}, \ and\
  \bibinfo {author} {\bibfnamefont {O.}~\bibnamefont {Zanusso}},\ }\href
  {\doibase 10.1103/PhysRevD.96.081701} {\bibfield  {journal} {\bibinfo
  {journal} {Phys. Rev.}\ }\textbf {\bibinfo {volume} {D96}},\ \bibinfo {pages}
  {081701} (\bibinfo {year} {2017})},\ \Eprint
  {http://arxiv.org/abs/1706.06887} {arXiv:1706.06887 [hep-th]} \BibitemShut
  {NoStop}%
\end{thebibliography}%

\end{document}